\documentclass{article}

\usepackage[all]{xy}
\usepackage{graphicx}

\usepackage{amsmath, amssymb, amsthm}

\newcommand{\calF}{\mathcal{F}}
\newcommand{\calI}{\mathcal{I}}
\newcommand{\calM}{\mathcal{M}}
\newcommand{\calP}{\mathcal{P}}

\newcommand{\bfF}{\mathbb{F}}
\newcommand{\bfG}{\mathbb{G}}
\newcommand{\bfR}{\mathbb{R}}
\newcommand{\bfZ}{\mathbb{Z}}
\newcommand{\bfX}{\mathbb{X}}

\newcommand{\id}{\mathrm{id}}

\newcommand{\ddt}{\frac{\d}{\d t}}

\newcommand{\be}{\begin{equation}}
\newcommand{\ee}{\end{equation}}

\newtheoremstyle{jvk-thm} %
  {}{}{\itshape}{}{\bfseries}{.}{ }{}
\newtheoremstyle{jvk-rem} %
  {}{}{\upshape}{}{\bfseries}{.}{ }{}

\theoremstyle{jvk-thm}
\newtheorem{definition}{Definition}[section]

\newtheorem{theorem}[definition]{Theorem}
\newtheorem{prop}[definition]{Proposition}
\newtheorem{corollary}[definition]{Corollary}

\newenvironment{mproof}{\textbf{Proof:}\,}{\hfill$\Box$}

\theoremstyle{jvk-rem}
\newtheorem{example}[definition]{Example}
\newtheorem{remarkth}[definition]{Remark}
\newenvironment{remark}{\begin{remarkth}}{\hfill$\diamond$\end{remarkth}}

\newenvironment{RomanList}{%
  \begin{enumerate}}{\end{enumerate}}

\renewcommand{\d}{\mathrm{d}}

\newcommand{\Sec}{\mathrm{Sec}}

\newcommand{\qfor}{\quad \text{for }}
\newcommand{\qforall}{\quad \text{for all}\:}

\newcommand{\where}{\quad \mathrm{where}\:}
\newcommand{\qqand}{\quad \mathrm{and} \quad}

\newcommand{\prl}[1][k]{P^{#1} \bfG}
\newcommand{\LAproj}{\mathfrak{P}}
\newcommand{\leg}{\bfF L}

\title{Discrete Lagrangian field theories on Lie groupoids} \author{Joris
  Vankerschaver\thanks{Research Assistant of the Research
    Foundation--Flanders (FWO-Vlaanderen)}, Frans Cantrijn \\
  {\small Department of Mathematical Physics and Astronomy, Ghent
    University } \\
  {\small Krijgslaan 281, B-9000 Ghent, Belgium } \\
  {\small e-mail: \{Joris.Vankerschaver, Frans.Cantrijn\}@UGent.be }}

\begin{document}
\maketitle

\begin{abstract}
  We present a geometric framework for discrete classical field
  theories, where fields are modeled as ``morphisms'' defined on a
  discrete grid in the base space, and take values in a Lie groupoid.
  We describe the basic geometric setup and derive the field equations
  from a variational principle.  We also show that the solutions of
  these equations are multisymplectic in the sense of Bridges and
  Marsden.  The groupoid framework employed here allows us to recover
  not only some previously known results on discrete multisymplectic
  field theories, but also to derive a number of new results, most
  notably a notion of discrete Lie-Poisson equations and discrete
  reduction.  In a final section, we establish the connection with
  discrete differential geometry and gauge theories on a lattice.
\end{abstract}

\section{Introduction}

The idea of studying mechanical systems on Lie groupoids first arose
in the context of discrete dynamical systems when Moser and Veselov
(see \cite{MoserVeselov91}) considered the pair groupoid $Q \times Q$
as a discretization of the tangent bundle $TQ$ and used it in their
study of discrete integrable systems.  Their idea was subsequently
used by Weinstein \cite{Weinstein96}, who introduced (among other
things) Lagrangian mechanics on an arbitrary Lie groupoid, established
a suitable variational principle for it and laid the foundations of
discrete reduction.

The theme of mechanics on a Lie groupoid was then picked up again in
\cite{GroupoidMech05}, in which the authors extended Weinstein's
approach by fully exploring the geometry of the various prolongation
bundles associated to the groupoid.  They gave a direct construction
of the Poincar\'e-Cartan forms and the Legendre transformations,
proved the symplecticity of the discrete flow and made the connection
with numerous examples of discrete mechanical systems that had been
studied before (see \cite{BobenkoSuris99CMP,
  MarsdenPekarskyShkoller99, Weinstein96} and the references therein).

Meanwhile, the foundational idea of Moser and Veselov of replacing
$TQ$ by the discretization $Q \times Q$, was extended to the case of
\emph{field theories} by Marsden, Patrick and Shkoller in
\cite{MPS98}.  Their objective was a systematic study of the geometry
of discrete multisymplectic field theories, aimed at the design of
robust numerical integrators that conserve an appropriate notion of
``symplecticity''.  The symplectic nature of their discrete field
theories is a consequence of the variational structure and is
expressed in terms of a set of distinct one-forms $\theta_L^i$, called
Poincar\'e-Cartan forms, such that $\sum_i \theta_L^i = \d L$, and
they observed that symplectic discretization schemes indeed yield
superior results.  

A similar approach, but aimed instead at Hamiltonian multisymplectic
PDEs, was proposed in \cite{Bridges01}, based on Bridges' notion of
multisymplecticity in \cite{Bridges97, BridgesDerks} (see also
\cite{Bridges05, LeimkuhlerReich}), and again it was observed that
multisymplectic discretizations indeed have remarkable energy and
momentum conservation properties.  Moreover, they showed that a number
of classical numerical schemes such as the Euler or Preissman box
scheme have a natural interpretation as multisymplectic integrators.

The objective of this paper is to establish the discrete counterpart
of Lagrangian multisymplectic field theory, in the case where the
discrete fields take values in an arbitrary Lie groupoid.  In doing
so, we extend both discrete Lagrangian field theory, as treated in
\cite{MPS98}, as well as mechanics on Lie groupoids
\cite{GroupoidMech05}.  We use techniques from groupoid mechanics and
show how they can be generalized quite easily to field theories.  In
doing so, we develop some new insights into some of the constructions
in \cite{MPS98}.  Finally, we present a number of new results which
require the full machinery developed here.  The most notable example
is a discrete version of the Lie-Poisson equations for field
theories.  We finish by presenting some remarks on discrete
differential geometry, as it turns out that our way of
modeling discrete fields is reminiscent of the way in which discrete
connections are usually introduced.

\section{Discrete mechanics on Lie groupoids} \label{sec:grpmech}

In this section, we recall some of the basic definitions and results
from the theory of Lie groupoids and algebroids.  It is not our
intention to give a detailed introduction to the suject: for a more
in-depth overview, the reader is referred to \cite{Mackenzie} and the
references therein.  We will also recall some of the constructions in
\cite{GroupoidMech05} that will be generalized in the next sections.
We note that the definition of a groupoid used here agrees with
\cite{GroupoidMech05, Weinstein96} but differs from \cite{Saunders04}
with respect to the order of writing the product $gh$.

\subsection{Lie groupoids}

A \emph{groupoid} is a set $G$ with a partial multiplication $m$, a
subset $Q$ of $G$ whose elements are called \emph{identities}, two
surjective maps $\alpha, \beta: G \rightarrow Q$ (called \emph{source}
and \emph{target} maps respectively), which both equal the identity
on $Q$, and an inversion mapping $i: G \rightarrow G$.  A
pair $(g,h)$ is said to be \emph{composable} if the multiplication
$m(g, h)$ is defined; the set of composable pairs will be denoted by
$G_2$.  We will denote the multiplication $m(g,h)$ by $gh$ and the
inversion $i(g)$ by $g^{-1}$.  In addition, these data must satisfy
the following properties, for all $g, h, k \in G$:
\begin{enumerate}
\item the pair $(g, h)$ is composable if and only if $\beta(g) =
  \alpha(h)$, and then $\alpha(gh) = \alpha(g)$ and $\beta(gh) =
  \beta(h)$;

\item if either $(gh)k$ or $g(hk)$ exists, then both do, and they are
  equal;

\item $\alpha(g)$ and $\beta(g)$ satisfy $\alpha(g)g = g$ and
  $g\beta(g) = g$;

\item the inversion satisfies $g^{-1}g = \beta(g)$ and $gg^{-1} =
  \alpha(g)$.
\end{enumerate}

On a groupoid, we have a natural notion of left translation $l_g$,
defined as $l_g(h) = gh$, for any $h \in G$ such that $\alpha(h) =
\beta(g)$.  There is a similar definition for a right translation
$r_g$.

A \emph{morphism of groupoids} is a pair of maps $\phi: G \rightarrow
G'$ and $f: Q \rightarrow Q'$ satisfying $\alpha' \circ \phi = f \circ
\alpha$, $\beta' \circ \phi = f \circ \beta$ and such that $\phi(gh) =
\phi(g)\phi(h)$ whenever $(g, h)$ is composable.  Note that $(\phi(g),
\phi(h))$ is a composable pair whenever $(g,h)$ is composable.

A \emph{Lie groupoid} is a groupoid for which $G$ and $Q$ are
differentiable manifolds, with $Q$ a closed submanifold of $G$, the
maps $\alpha, \beta, m$ and $i$ are smooth and $\alpha$ and $\beta$
are submersions.  We denote by $\calF^\alpha(g)$ the $\alpha$-fibre
through $g \in G$, i.e. $\calF^\alpha(g) = \alpha^{-1}(\alpha(g))$,
with a similar definition for $\calF^\beta(g)$.  As $\alpha$ and
$\beta$ are submersions, both $\calF^\alpha(g)$ and $\calF^\beta(g)$
are closed submanifolds of $G$.

Any Lie group $G$ can be considered as a Lie groupoid over a
singleton $\{e\}$, where the anchors $\alpha, \beta$ map any
element onto $x$ and the multiplication is defined everywhere.
Another example of a Lie groupoid is the pair groupoid $Q \times Q$,
where $\alpha(q_1, q_2) = q_1$, $\beta(q_1, q_2) = q_2$, and
multiplication is defined as $(q_1, q_2) \cdot (q_2, q_3) = (q_1,
q_3)$.  For other, less trivial examples, we refer to the works
mentioned above.

\subsection{Lie algebroids}

A \emph{Lie algebroid} over $Q$ is a vector bundle $\tau: E
\rightarrow Q$ together with a vector bundle map $\rho : E
\rightarrow TQ$ (called the \emph{anchor map} of the Lie algebroid)
and a bracket $[\cdot, \cdot] : \Sec(E) \times \Sec(E) \rightarrow
\Sec(E)$ defined on the sections of $\tau$, such that
\begin{enumerate}
  \item $\Sec(E)$ is a real Lie algebra with respect to $[\cdot,
    \cdot]$; 
  \item $\rho([\phi, \psi]) = [\rho(\phi), \rho(\psi)]$, for all
    $\phi, \psi \in \Sec(E)$, where the bracket on the right-hand side
    is the usual Lie bracket of vector fields on $Q$ and we write the
    composition $\rho \circ \phi$ as $\rho(\phi)$;
  \item $[\phi, f \psi] = f [\phi, \psi] + \rho(\phi)(f) \psi$,
    for all $\phi, \psi \in \Sec(E)$ and $f \in C^\infty(Q)$.
\end{enumerate}

The Lie algebroid structure allows us to define an exterior
differential $\d_E$ on the space of sections of $\bigwedge^\ast
(E^\ast)$, as follows: for functions $f \in C^\infty(Q)$, we put $\d_E
f(v) = \rho(v)f$, for $v \in E$, while for sections $\theta$ of
$\bigwedge^k(E^\ast)$, we define $\d_E \theta$ by
\begin{multline*}
      \d_E \theta(v_0, v_1, \ldots, v_k) = \sum_i \rho(v_i)
      \theta(v_0, \ldots, \hat{v}_i, \ldots, v_k)  \\
      + \sum_{i < j} (-1)^{i+j} \theta([v_i,v_j], v_0, \ldots,
      \hat{v}_i, \ldots, \hat{v}_j, \ldots, v_k).
\end{multline*}
It can be shown that $\d_E$ is nilpotent: $\d_E^2 = 0$.

To any Lie groupoid $G$ over $Q$ one can associate a Lie algebroid
$\tau: AG \rightarrow Q$ as follows.  At each point $x \in Q$, the
fibre $A_xG$ is the vector space $V_x \alpha = \ker T_x \alpha$ and
the anchor map $\rho$ on $A_xG$ is identified with the restriction of
$T_x \beta$ to $V_x \alpha$.  In order to define the bracket on the
space of sections, we note that there exists a bijection between
sections of $\tau$ and left- and right-invariant vector fields on $G$.
More specifically, if $v$ is a section of $\tau$, then the left- and
right-invariant vector fields are denoted as $v^L$ and $v^R$
respectively, and defined by \be \label{vectorfields} v^L(g) =
T_{\beta(g)}l_g(v_{\beta(g)}) \qqand v^R(g) = T_{\alpha(g)}(r_g \circ
i)(v_{\alpha(g)}).  \ee Let $v$ and $w$ be sections of $\tau$.  The
bracket $[v, w]$ is then defined by noting that $[v^L, w^L]$ is again
a left-invariant vector field, and putting
\[
  [v, w]^L = [v^L, w^L].
\]
We remark that our definition of $v^R$ differs in sign from the one
used in \cite{GroupoidMech05}.

Conversely, we say that a Lie algebroid $\tau: E \rightarrow Q$ is
\emph{integrable} whenever one can find a Lie groupoid such that $E$
is its associated Lie algebroid.  It has been known for some years
that not all Lie algebroids are integrable.  Necessary and sufficient
conditions for integrability have been given in \cite{Fernandes}.    

The Lie algebroid of a Lie group $G$ is just its Lie algebra.  The Lie
algebroid of the pair groupoid $Q \times Q$ is the tangent bundle
$TQ$. 

\begin{remark} \label{remark:notation} For a given section $v$ of
  $\tau$, we have denoted the corresponding left- and right-invariant
  vector fields as $v^L$ and $v^R$, respectively.  We will also use
  this notation for the pointwise operation, by denoting, for $v_x$ an
  element of $A_x G$ and $g \in \alpha^{-1}(x) \subset G$, the left
  translated vector $T_x l_g (v_x)$ as $(v_x)^L(g)$, and similarly the
  right translated vector $T_x(r_g \circ i)(v_x)$ as $(v_x)^R(g)$.
\end{remark}

\subsection{Lie algebroid morphisms}

Consider two vector bundles $\tau': E' \rightarrow Q'$ and $\tau: E
\rightarrow Q$, and let $(\Phi, \varphi)$ be a vector bundle map from
$\tau'$ to $\tau$.  Let $\theta$ be a section of $\bigwedge^k(E^\ast)$.
Then the \emph{pullback of $\theta$ by $(\Phi, \varphi)$} is the
section $\Phi^\star \theta$ of $\bigwedge^k(E'^\ast)$ defined as
\[
  (\Phi^\star \theta)_q(v_1, \ldots, v_k) =
  \theta_{\varphi(q)}(\Phi(v_1), \ldots, \Phi(v_k)), \quad v_1,
  \ldots, v_k \in E_q.
\]

Note that we used a ``star'' $\star$ instead of an ``asterisk'' $\ast$
to denote the pullback, which should serve as a reminder that we
consider the pullback of $\theta$ by a \emph{bundle map} rather than
by an arbitrary differentiable map from $E'$ to $E$.

Now, assume that both $\tau$ and $\tau'$ are equipped with the
structure of a Lie algebroid over $Q$.  In this case, a vector bundle
map $(\Phi, \varphi)$ is said to be a \emph{morphism of Lie
  algebroids} if for each section $\theta$ of $\bigwedge^k(E^\ast)$,
\[
    \Phi^\star \d_{E} \theta = \d_{E'} \Phi^\star \theta,
\]
where $\d_{E}$ and $\d_{E'}$ are the differentials on $E$ and $E'$,
respectively.  In other words, $(\Phi, \varphi)$ is a \emph{chain
  map}.  In \cite{Higgins90, Martinez04, Martinez05}, a number of
equivalent conditions are investigated for a bundle map to be a
morphism of Lie algebroids.

\subsection{The prolongation of a Lie groupoid over a
  fibration} \label{sec:prlgroupoid} 

Let $G$ be a Lie groupoid over a manifold $Q$ with source and target
maps $\alpha$ and $\beta$ and consider a fibration $\pi: P \rightarrow
Q$.  The \emph{prolongation} $P^\pi G$ is the Lie groupoid over $P$
defined as
\[
  P^\pi G = \{ (g; p_1, p_2) \in G \times P \times P: \pi(p_1) =
  \alpha(g) \text{ and } \beta(g) = \pi(p_2) \}.
\]
Alternatively, $P^\pi G$ is defined by means of the following
commutative diagram:
\be \label{diag:prlgroupoid} \xymatrix{
  P^\pi G \ar[r] \ar[d] & P \times P \ar[d]^{\pi \times \pi} \\
  G \ar[r]_{\alpha \times \beta} & Q \times Q
}\ee

It can be shown that $P^\pi G$ is a Lie groupoid over $P$, with source
and target mappings $\alpha^\pi, \beta^\pi: P^\pi G \rightarrow P$
defined as
\[
  \alpha^\pi(g; p_1, p_2) = p_1 \qqand \beta^\pi(g; p_1, p_2) = p_2,
\]
and with multiplication given by
\[
  (g; p_1, p_2) (h; p_2, p_3) = (gh; p_1, p_3).
\]
Note that $\alpha^\pi(h; p_2, p_3) = \beta^\pi(g; p_1, p_2)$ implies
that $\alpha(h) = \beta(g)$.  Finally, the inversion mapping is
defined as 
\[
i: (g; p_1, p_2) \mapsto (g^{-1}; p_2, p_1),
\]
and we can regard $P$ as a subset of $P^\pi G$ via the identification
$p \mapsto (\pi(p); p, p)$.

\subsubsection{The prolongation $PG$} \label{sec:prlPG}

There is one particular prolongation that will play a significant role
in what follows.  It is obtained by taking for the fibration $\pi: P
\rightarrow Q$ in (\ref{diag:prlgroupoid}) the Lie algebroid projection
$\tau: AG \rightarrow Q$ to obtain
\[
  P^\tau G \subset G \times AG \times AG
\]
which, henceforth, we also simply denote as $PG$.  We recall that $PG$
consists of triples $(g; v_x, w_y)$, where $g \in G$, $v_x \in A_x G$,
$w_y \in A_y G$, and $x = \alpha(g)$, $y = \beta(g)$.  It is pointed
out in \cite{GroupoidMech05, Saunders04} that $PG$ is isomorphic as a
vector bundle over $G$ to
the direct sum $V \beta \oplus V\alpha$, where $V\alpha$ is the
subbundle of $TG$ consisting of $\alpha$-vertical vectors (and
similarly for $V\beta$); the isomorphism $\Theta: PG \rightarrow
V\beta \oplus V\alpha$ is defined by
\be \label{isomorphism} 
  \Theta(g; u_{\alpha(g)}, v_{\beta(g)}) = 
    ( T(r_g \circ i)(u_{\alpha(g)}), Tl_g(v_{\beta(g)})).  
\ee 
It should also be remarked that $PG$ is a vector bundle over $G$, and
in fact, $PG$ can be endowed with the structure of an integrable Lie
algebroid over $G$, where the anchor map $\hat{\rho}: PG \rightarrow
TG$ is given by
\[
  \hat{\rho}: (g; u_{\alpha(g)}, v_{\beta(g)}) \mapsto T(r_g \circ
  i)(u_{\alpha(g)}) + Tl_g(v_{\beta(g)}) = (u_{\alpha(g)})^R(g) +
  (v_{\beta(g)})^L(g).  
\]
Given a pair of sections $u, v$ of $AG$, one can construct a section
of $PG \rightarrow G$, shortly denoted by $(u, v)$, by considering the
map $g \mapsto (g; u_{\alpha(g)}, v_{\beta(g)})$.  The Lie bracket of
sections of $PG$ is then determined by the following definition:
\[
  [ (u, v), (u', v')]_{PG} = ( [u, u'], [v, v'] ),
\]
where $u, u', v, v'$ are sections of $AG$ (see
\cite[Thm.~3.1]{GroupoidMech05}).

\subsection{The prolongation of a Lie algebroid over a
  fibration} \label{sec:prlalg} 

Let $\tau: E \rightarrow Q$ be a Lie algebroid and consider a
fibration $\pi: P \rightarrow Q$.  The \emph{prolongation} $P^\pi E$
is the Lie algebroid over $P$ defined as 
\[
  P^\pi E = \{ (a, v) \in E \times TP: \rho(a) = T\pi(v) \},
\]
or by the following commutative diagram as
\be\xymatrix{
  P^\pi E \ar[d] \ar[r] & TP \ar[d]^{T\pi} \\
  E \ar[r]_{\rho} & TQ
} \label{diag:prlalg}
\ee

We denote by $\hat{\pi}: P^\pi E \rightarrow P$ the map defined as
$\hat{\pi}(a, v) = \pi_{TP}(v)$, where $\pi_{TP}: TP \rightarrow P$ is
the tangent bundle projection of $P$.  It can be shown that
$\hat{\pi}: P^\pi E \rightarrow P$ can be given the structure of a Lie
algebroid (see \cite{Higgins90, Martinez01, Saunders04}).

\subsubsection{The prolongations $P^\tau(AG)$ and
  $P^{\tau^\ast}(AG)$} \label{sec:prltangent}

Let $G$ be a Lie groupoid over a manifold $Q$ with Lie algebroid
$\tau: AG \rightarrow Q$.  By taking for the fibration $\pi$
underlying diagram (\ref{diag:prlalg}) the map $\tau$, we
obtain the prolongation $P^\tau(AG)$.  It is very useful to think of
$P^\tau(AG)$ as a sort of Lie algebroid analogue of the \emph{tangent
  bundle} to $AG$.  Indeed, $P^\tau(AG)$ can be equipped with 
geometric objects, such as a Liouville section and a vertical
endomorphism, which have their counterpart in tangent bundle geometry.

Similarly, by taking for $\pi: P \rightarrow Q$ the dual bundle
$\tau^\ast: A^\ast G \rightarrow Q$, we obtain the prolongation
$P^{\tau^\ast}(AG)$, which is a Lie algebroid over $A^\ast G$ and
should be thought of as the Lie algebroid analogue of the
tangent bundle to $A^\ast G$.  Just as any cotangent bundle is
equipped with a canonical one-form, there exists a canonical section
\[
  \theta : A^\ast G \rightarrow \left[ P^{\tau^\ast}(AG) \right]^\ast, 
\]
defined as follows: for $\alpha \in A^\ast G$ and $(v, X_\alpha) \in
(P^{\tau^\ast}(AG))_\alpha$, we put $\theta_\alpha(v, X_\alpha) =
\alpha(v)$. In the case that $G$ is the pair groupoid $Q \times Q$, we
have that $A^\ast G = T^\ast Q$ and we obtain the usual canonical
one-form on $T^\ast Q$.

It was shown in \cite{Higgins90} that $P^\tau(AG)$, the prolongation
of the Lie algebroid $AG$, is isomorphic to $A(PG)$, the Lie algebroid
associated to the prolongation Lie groupoid $PG$.

\subsubsection{The prolongations $P^\alpha(AG)$ and
  $P^\beta(AG)$} \label{sec:prlalphabeta} 

Associated to the source and target mappings $\alpha$ and $\beta$ of a groupoid
$G$ there are two prolongations $P^\alpha(AG)$ and $P^\beta(AG)$,
whose fibres over $G$ are defined as follows: for each $g \in G$, put
\[
  P^\alpha_g(AG) = \{ (v_{\alpha(g)}, X_g) \in A_{\alpha(g)}G \times
  T_g G : T\tau(v_{\alpha(g)}) = T\alpha(X_g) \}
\]
and
\[
  P^\beta_g(AG) = \{ (v_{\beta(g)}, X_g) \in A_{\beta(g)}G \times
  T_gG: T\tau(v_{\beta(g)}) = T\beta(X_g) \}.
\]
Both of these algebroids are integrable: indeed, it follows from the
general theory that $P^\alpha(AG)$ is isomorphic to the Lie algebroid
of the prolongation $P^\alpha G$, and similarly for $P^\beta(AG)$.

Furthermore, we remark that there are two distinguished mappings from
$PG$ (regarded as a Lie \emph{algebroid} over $G$) into $P^\alpha(AG)$
and $P^\beta(AG)$, given by
\[
  A(\Phi^\alpha) : (u_{\alpha(g)}, g, v_{\beta(g)}) \mapsto
  (u_{\alpha(g)}, T(r_g \circ i)(u_{\alpha(g)}) + Tl_g(v_{\beta(g)}))
  \in P^\alpha(AG)
\]
and
\[
  A(\Phi^\beta) : (u_{\alpha(g)}, g, v_{\beta(g)}) \mapsto
  (v_{\beta(g)}, T(r_g \circ i)(u_{\alpha(g)}) + Tl_g(v_{\beta(g)}))
  \in P^\beta(AG).
\]
The notations $A(\Phi^\alpha)$ and $A(\Phi^\beta)$ serve as a reminder
of the fact that these Lie algebroid maps stem from morphisms between
the corresponding groupoids (see \cite{GroupoidMech05}).

\section{The discrete jet bundle.  Discrete
  fields} \label{sec:discretebundle} 

Let us now turn to field theory.  As is customary in most geometric
treatments, we model physical fields as sections of a fibre bundle
$\pi: Y \rightarrow X$.  This approach has received a lot of attention
in the past and we refer to \cite{CrampinMS, GimmsyI, Saunders89} for
more information.  For the sake of simplicity, we will assume from now
on that the base space $X$ of $\pi$ is $\bfR^2$, and that $\pi$ is
trivial, i.e. $\pi$ is given by $\pi: \bfR^2 \times Q \rightarrow
\bfR^2$, where $Q$ is the standard fibre.

It is our aim in this section to present a geometric approach to
\emph{discrete} field theories.  A crucial element of this setup is
the concept of \emph{discrete jet bundle}.  Before going into details,
it is perhaps useful to start with a quick overview of what our
construction entails.

\subsection{Overview}

We will introduce a notion of ``discrete jet bundle of $\pi$'', using two
essentially different ingredients:

\begin{enumerate}
\item The existence of a \emph{mesh} in $X = \bfR^2$, consisting of a
  discrete subset $V$ of $X$, whose elements are called
  \emph{vertices}, and a set $E$ of \emph{edges}, which are line
  segments between pairs of vertices.  Associated to such a mesh is a set
  of faces, where a face $f$ is a region in $\bfR^2$ bounded by edges,
  and such that there are no edges in the interior of $f$.

\item A groupoid $G$ over the standard fibre $Q$ of $\pi$.  This is a
  new element, and its role will become clear in a moment.
\end{enumerate}

We will define a discrete jet as a mapping which assigns to each edge
of the mesh
an element of $G$ such that two edges which have a vertex in common
are mapped onto composable elements of $G$.  We will show that each such
mapping gives rise to a groupoid morphism from the pair groupoid $V
\times V$ (where $V$ is the set of vertices) to $G$.  In
section~\ref{sec:pair} we will treat the particular case where $G$ is
the pair groupoid $Q \times Q$.  In that case, a discrete field is an
assignment of an element of $Q$ to each point of a grid in $X$, which
is a natural way, used for example in finite-difference methods, to
think of discrete fields (see \cite{MPS98}).

\paragraph{The manifold $\bfG^k$}

There is another, equivalent, way of thinking of discrete jets, which
is closely related to the way in which continuous jets are
interpreted.  Recall that we considered a trivial bundle $\pi: \bfR^2
\times Q \rightarrow \bfR^2$.  In this case, the jet bundle $J^1\pi$
is isomorphic to the product space $\bfR^2 \times J^1_0(\bfR^2, Q)$,
where $J^1_0(\bfR^2, Q)$ is the manifold of $1$-jets at $0$ of maps
$\varphi: \bfR^2 \rightarrow Q$, which is itself isomorphic to the
Whitney sum $TQ \oplus TQ$.  Incidentally, this is the starting point
for the so-called $k$-symplectic (here $k = 2$) treatment of field
theories (see \cite{Gunther, Salgado04} and the references therein).
Hence, a natural interpretation of a jet at a point $x \in \bfR^2$ is
as an element of $TQ \oplus TQ$.

Let us now repeat this procedure for the discrete case.  We start from
the base space $X = \bfR^2$ and a given mesh $(V, E)$.  As we argued
before, there is a natural definition of the set of faces of this mesh
as (connected) regions of the plane bounded by edges.  Furthermore, as
the edges are represented by pairs of vertices, and faces are defined
by specifying their bounding edges, a face is completely determined by
its corner vertices $x_1, \ldots, x_k$, where the vertices are ordered
in such a way that the bounding edges are $(x_i, x_{i+1})$ (for $i =
1, \ldots, k-1$) and $(x_1, x_k)$.  Each of the pairs $(x_i, x_{i+1})$
is a Veselov-type discretization of a tangent vector and, hence a face
is a natural way of representing a set of $k - 1$ vectors.  As soon as
$k > 3$, this set can never be linearly independent.  However, it turns
out that this makes essentially no difference for the discrete
approach, and might even have certain benefits in the design of
numerical methods (see \cite[p.~42]{MPS98}).  We will consider in
general only meshes in which each face has the same number of edges,
which we denote henceforth as $k$.

Recall that in the continuous case, we interpreted jets as elements of
$TQ \oplus TQ$ by considering the values they take on the standard
basis of $\bfR^2$.  Let us now define a discrete jet as an assignment
of $k$ points in $Q$ to any face $\{x_1, \ldots, x_k\}$ of the mesh,
in other words: a $k$-tuple $\{q_1, \ldots, q_k\}$ of points in $Q$
(together with the face $\{x_1, \ldots, x_k\}$).  Hence the fibre part
of our space (the part involving only $Q$) of jets is really a
discretization of $TQ \oplus \cdots \oplus TQ$ ($k$ times).

As a slight generalization, we can easily replace the pair groupoid $Q
\times Q$ by an arbitrary groupoid $G$ over $Q$: in this case, we are
led to the study of a similar manifold $\bfG^k$ (consisting of
``faces'' in $G$, to be specified later), which is the discrete
counterpart of $AG \oplus \cdots \oplus AG$.

In proposition~\ref{prop:onetoone}, we will show how both
points of view, i.e. discrete jets on the one hand and the manifold
$\bfG^k$ on the other hand, are related.

\subsection{Discretizing the base space}

\subsubsection{The mesh} \label{sec:mesh}

To discretize $X = \bfR^2$, we will use the concept of a \emph{mesh}
embedded in $X$.  Intuitively, such a mesh consists of a discrete
subset $V$ of $X$ together with a number of relations specifying which
points of $V$ ``belong together''.  This can be made more rigourous by
means of some elementary concepts from graph theory, which we now
review.

A \emph{graph} is a pair of sets $(V, E)$ such that $E$ is a subset of
$V \times V$.  In contrast to what is usually assumed in graph theory,
we will allow $V$ and $E$ to be (countably) infinite.  The elements of
$V$ are called \emph{vertices}, while those of $E$ are called
\emph{edges}.  Note that the edges in $E$ are \emph{undirected}.

A graph is \emph{simple} if there is at most one edge connecting each
pair of distinct vertices.  In this case, let us represent an edge $e
\in E$ by its incident vertices as $e \rightsquigarrow \{x, y\}$.  A
\emph{path} between two vertices $x$ and $y$ is a sequence of edges
$\{x, p_1\}, \{p_1, p_2\}, \ldots, \{p_l, y\}$.  A graph is said to be
\emph{connected} if there exists a path between any two vertices.  In
the sequel, we will only consider connected, simple graphs, with the
additional condition that there are no ``loops'', i.e. no edges $e$
whose incident vertices coincide.

A \emph{planar graph} is a graph $(V, E)$ where $V$ is a subset of
$\bfR^2$ and the edges are curves in $E$ connecting pairs of vertices
such that if any two edges intersect, they do so in a common vertex.
For a planar graph, there is a notion of \emph{face}, defined as
follows.  Consider the \emph{geometric realisation} $|E|$ of $(V, E)$,
which is just the union of all edges.  The complement $\bfR^2
\backslash |E|$ of $|E|$ is a disconnected set, whose connected
components are the \emph{faces} of the planar graph $(V, E)$.  A face
is therefore a region in the plane, bounded by a number of edges.

The \emph{degree of a face} is defined as the number of edges that
make up the boundary of that face.  Dually, the \emph{degree of a
  vertex} is defined as the number of edges arriving in that vertex.

\begin{figure}
\begin{center}
  \includegraphics{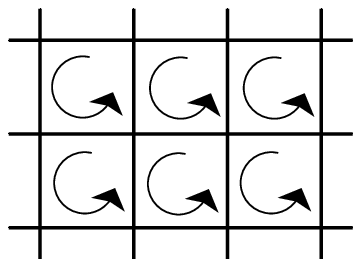}
  \caption{Square mesh in $\bfR^2$, with counterclockwise
    orientation.}
  \label{fig:quadrangles}
\end{center}
\end{figure}

\begin{definition} \label{def:mesh} A \emph{mesh} in $X = \bfR^2$ is a
  simple and connected planar graph $(V, E)$ in $X$ such that
  the following conditions are satisfied:
  \begin{enumerate}
    \item the edges are realised as segments of straight lines in
      $\bfR^2$; 
    \item the degree of the faces is constant and equal to some
      natural number $k > 2$;
    \item the degree of the vertices is always larger than two.
  \end{enumerate}
\end{definition}
It has to be stressed that the nature of this graph is left entirely
unspecified and should be dictated by the problem under scrutiny.
Throughout this text, we will illustrate our theory from time to time
using some elementary meshes, of which the covering of $\bfR^2$ by
quadrangles, as in figure~\ref{fig:quadrangles}, is the most
straightforward.  This mesh was also used in \cite{MPS98}.

A few remarks concerning the above definition are in order.  The fact
that, given a mesh $(V, E)$, the elements of $E$ are realised as
straight line segments, implies that each edge is determined by its
begin and end vertex.  Similarly, a face $f$ is determined by its $k$
bounding edges $e_1, \ldots, e_k$, each of which can be represented as
a pair of vertices $e_i = \{x_i, x_{i+1}\}$ (where $x_{k+1} = x_1$),
and, hence, $f$ is determined by specifying the set of its ``corner''
vertices:
\[
  f \rightsquigarrow \{x_1, \ldots, x_k\}.
\]
The set of all faces associated to a mesh $(V, E)$ will be denoted by
$F$.  One can envisage a more general situation in which the edges are
allowed to be more general curves.

\begin{remark}
  In a recent paper \cite{Wise} on lattice gauge theories, the author
  introduces a discretization of space-time by means of a hypothetical
  ``$n$-graph'' structure, which is a list of data $X_0, X_1, X_2,
  \ldots$, where $X_0$ is a set of vertices, $X_1$ a set of edges, and
  so on, with sets $X_i$ of higher-dimensional objects.  These sets
  have to specify various incidence relations, the nature of which is
  still not entirely clear.  However, the concepts of $n$-graphs or
  $n$-complexes (weaker versions of $n$-graphs) would be useful in
  generalising our theory to the case where the base space is no
  longer two-dimensional or Euclidian.
\end{remark}

\subsubsection{The local groupoid $E$} \label{sec:localgroupoid}

In order to bring to the fore the algebraic character of the set of
edges $E$ of a given mesh $(V, E)$, we construct a new set $E'$, whose
elements are \emph{ordered} pairs $(x, y) \in V \times V$ satisfying
the following axioms:
\begin{enumerate}
  \item $(x, x) \in E'$ for all $x \in V$;
  \item if $\{x, y\}$ is an element of $E$, then $(x, y) \in E'$ and
    $(y, x) \in E'$.
\end{enumerate}
The important difference between $E$ and $E'$ is that the elements of
$E$ are \emph{undirected} edges, whereas the elements of $E'$ are
\emph{directed}.  As we will no longer have a use for $E$, no
confusion can arise if we, henceforth, denote $E'$ simply by $E$.

If we define the source and target mappings $\alpha_X, \beta_X: E
\rightarrow V$ in the usual way as $\alpha_X(x, y) = x$ and
$\beta_X(x, y) = y$, then $E$ is a subset of the pair groupoid $V
\times V$, satisfying all but one of the axioms of a discrete
groupoid: if $e_1 = (x, y)$ and $e_2 = (y, z)$ are elements of $E$
such that $\beta_X(e_1) = \alpha_X(e_2)$, then the multiplication
$e_1\cdot e_2$, defined as $e_1 \cdot e_2 = (x, z)$, is an element of
$V \times V$ but not necessarily of $E$.

This is strongly reminiscent of the concept of \emph{local groupoid}
introduced by Van Est in \cite{VanEst} in the context of Lie groupoids
as, roughly speaking, differentiable groupoids in which the condition
$\beta(e_1) = \alpha(e_2)$ is necessary but not sufficient for the product
$e_1 \cdot e_2$ to exist.  Even though in its original definition this
concept makes no sense for discrete spaces, the name is nevertheless
quite appropriate and so we will continue to refer to $E$ as a local
groupoid.

\subsubsection{The set of $k$-gons $\bfX^k$}

We now introduce the set of $k$-gons $\bfX^k$.  The elements of this
set are the faces of the mesh, but with a consistent orientation.
Indeed, the natural orientation of $X = \bfR^2$ allows us to write
down the edges of each face $f$ in (say) counterclockwise direction:
\[
  f = (x_k, x_1), (x_1, x_2), \ldots, (x_{k-1}, x_k).
\]
We now introduce $\bfX^k$ as the set of all faces, considered as
$k$-tuples of edges written down in the counterclockwise direction: 
\[
\bfX^k = \left\{ \big((x_k, x_1), (x_1, x_2), \ldots, (x_{k-1},
  x_k)\big) \where \{x_1, \ldots, x_k\} \in F \right\}.
\]
We will also refer to the elements of $\bfX^k$ as \emph{$k$-gons} and
denote them as $[x] := ((x_k, x_1), (x_1, x_2), \ldots, (x_{k-1},
x_k))$. To refer to the $i$th component of a $k$-gon $[x]$, we will
use the subscript notation: $[x]_1 = (x_k, x_1)$ and $[x]_i =
(x_{i-1}, x_i)$ for $i = 2, \ldots, k$.  In the following, we will
assume that the indices are defined ``modulo $k$, plus one'', which
allows us to write $[x]_i = (x_{i-1}, x_i)$, for all $i = 1, \ldots,
k$.

It is useful to note that a $k$-gon is not changed by a cyclic
permutation of its elements and that the common edge of two adjacent
$k$-gons is traversed in opposite directions.

\begin{example}
  In the example given in figure~\ref{fig:quadrangles}, the degree of
  each face is exactly four as each face is made up of four edges.
  The elements of $\bfX^4$ are the faces with the counterclockwise
  orientation indicated on the figure.
\end{example}

\subsection{The discrete jet space $\bfG^k$}

We now complete our programme of discretizing the jet bundle of $\pi$ by
constructing over the fibre $Q$ a structure $\bfG^k$ similar to
$\bfX^k$.  The elements of $\bfG^k$ are $k$-gons in $G$, each of which
is an approximation of a frame by $k$ groupoid elements.

\begin{definition}
  The \emph{discrete jet bundle} is the manifold $\bfG^k$ consisting
  of all ordered $k$-tuples $(g_1, \ldots, g_k) \in G \times \cdots
  \times G$ such that
  \[
    (g_1, g_2), (g_2, g_3), \ldots, (g_k, g_1) \in G_2 \qqand 
      g_1 \cdot g_2 \cdots g_k = \alpha(g_1) ( = \beta(g_k) ).
  \]
\end{definition}

Elements of $\bfG^k$ will be denoted as $[g] = (g_1, \ldots, g_k)$, and,
with the ``modulo'' convention introduced above, 
a subscript will be used to refer to the individual components: $[g]_i =
g_i$.  Note that, whereas $\bfX^k$ is a discrete set due to its
compatibility with the mesh, $\bfG^k$ is a smooth manifold and $\dim
\bfG^k \ge \dim G$.

The discrete jet bundle $\bfG^k$ can be equipped with the following two
operations: 
\begin{enumerate}
  \item the inverse of a given $k$-gon $[g]$, denoted as $[g]^{-1}$
    and defined as
    \[
      [g]^{-1} = (g_k^{-1}, g_{k-1}^{-1}, \ldots, g_1^{-1});
    \]

  \item a collection of $k$ mappings $\alpha^{(i)}: \bfG^k \rightarrow
    Q$, called \emph{generalized source maps} and defined as
    $\alpha^{(i)}([g]) = \alpha(g_i)$.
\end{enumerate}

\subsection{Discrete fields}

The idea of a ``discrete field'' can be expressed in terms of a mapping
that associates to each edge (i.e. to each element of the set $E$, in
the extended sense of subsection~\ref{sec:localgroupoid}) an element of
the groupoid $G$, and to each vertex in $V$ a unit of $G$, such that
whenever two edges are composable, so are their images.

\begin{definition} \label{def:DJ}
  A \emph{discrete field} is a pair $\phi = (\phi_{(0)}, \phi_{(1)})$,
  where $\phi_{(0)}$ is a map from $V$ to $Q$ and $\phi_{(1)}$ is a
  map from $E$ to $G$ such that 
  \begin{enumerate}
    \item $\alpha( \phi_{(1)}(x, y) ) = \phi_{(0)}(x)$ and $\beta(
      \phi_{(1)}(x, y) ) = \phi_{(0)}(y)$; \label{item:compat}

    \item for each $(x, y) \in E$, $\phi_{(1)}(y, x) = [\phi_{(1)}(x,
      y)]^{-1}$; \label{item:compose}
    \item for all $x \in V$, $\phi_{(1)}(x, x) = \phi_{(0)}(x)$.
  \label{item:unit}
  \end{enumerate}
\end{definition}

The definition we have given here is strongly reminiscent of that of a
groupoid morphism.  Of course $E$ is not a proper groupoid but just a
subset of $V \times V$.  However, a discrete field can easily be
extended to a groupoid morphism from $V \times V$ into $G$, as we now
show. 

\begin{prop} \label{prop:extension} Let $\phi = (\phi_{(0)},
  \phi_{(1)})$ be a discrete field.  Then there exists a unique
  groupoid morphism $(\varphi, f): V \times V \rightarrow G$ extending
  $\phi$.
\end{prop}
\begin{mproof}
First of all, we define $f(x) := \phi_{(0)}(x) \in Q$.  Now, let $(x, y)$ be any
element of $V \times V$.  If $(x, y) \in E$, then we put $\varphi(x,
y) := \phi_{(1)}(x, y)$.  If $(x, y) \notin E$, then, because of the
connectivity of the mesh (see definition~\ref{def:mesh}), there exists
a sequence $(x, u_1), (u_1, u_2), \ldots, (u_l, y)$ in $E$
such that in the pair groupoid $V \times V$, 
\be \label{seq} 
  (x, y) = (x, u_1) \cdot (u_1, u_2) \cdots (u_l, y).  
\ee 
We now put $\varphi(x, y) = \phi(x, u_1) \cdot \phi(u_1, u_2) \cdots
\phi(u_l, y)$.  As each factor on the right-hand side is composable
with the next (see property (\ref{item:compat}) in def.~\ref{def:DJ}), this
multiplication is well defined.  We only have to prove that
$\varphi(x, y)$ does not depend on the sequence used in (\ref{seq}).
Therefore, consider any other decomposition of $(x, y)$ as a product
in $V \times V$ of elements of $E$, i.e.
\be \label{prod} 
  (x, y) = (x, u'_1) \cdot (u'_1, u'_2) \cdots (u'_m, y).  
\ee 
and form the product
\[
  (x, x) = (x, u_1) \cdot (u_1, u_2) \cdots (u_l, y) \cdot (y, u'_m)
  \cdot (u'_m, u'_{m-1}) \cdots (u'_1, x).
\]
By acting on both sides with $\varphi$, we obtain 
\[
  f(x) = \varphi(x, u_1) \cdots \varphi(u_l, y) \cdot [\varphi(u_m',
  y)]^{-1} \cdots [\varphi(x, u_1')]^{-1}
\]
and therefore 
\[
  f(x) \varphi(x, u_1') \cdots \varphi(u_m', y) = \varphi(x, u_1)
  \cdots \varphi(u_l, y).
\]
By noting that $f(x) = \alpha( \varphi(x, u_1') )$, a left-sided unit,
we obtain the desired path independence.

To prove that $(\varphi, f)$ is unique, we consider a second groupoid
morphism $(\varphi', f')$ extending $\phi$, i.e. such that 
\[
  \varphi'(x, y) = \varphi(x, y) = \phi(x, y) \qfor (x,y) \in E. 
\]
Then, let $(x, y)$ be an arbitrary element of $V \times V$.  By
writing $(x, y)$ as a sequence of elements in $E$ as in (\ref{prod}),
and applying $\varphi'$ to this product, we may conclude that
$\varphi'$ coincides with $\varphi$ on the whole of $V \times V$.
\end{mproof}

\begin{remark}\footnote{We are grateful to R. Benito and D. 
    Mart\'{\i}n de Diego for pointing out to us this example as well 
    as the absence of property~\ref{item:unit} from 
    Definition~\ref{def:DJ} in an earlier version of this paper.}
  The preceding proposition makes clear why property~\ref{item:unit}
  of definition~\ref{def:DJ} cannot be omitted.  Indeed, consider the
  Lie group $G = GL(2, \bfR)$, and let $(\phi_{(0)}, \phi_{(1)})$ be
  the pair of constant maps defined as
  \[
    \phi_{(0)}(x) = \left(
        \begin{matrix} 1 & 0 \\ 0 & 1 \end{matrix}  \right)
      \qqand
    \phi_{(1)}(x, y) = \left( \begin{matrix} 0 & 1 \\ 1 & 0 
        \end{matrix} \right).
  \]
  The pair $(\phi_{(0)}, \phi_{(1)})$ satisfies the requirements of
  definition~\ref{def:DJ} except for property~\ref{item:unit}, but
  cannot be extended to a groupoid morphism.
\end{remark}

Henceforth, we will also write
$\phi$ for the unique morphism extending a given discrete field $\phi
= (\phi_{(0)}, \phi_{(1)})$.  

From a physical point of view, we are led to consider the mesh $(V,
E)$ in $X$
and hence the local groupoid $E$, and we define a discrete field to
attach a groupoid element to each element of $E$.  From a mathematical
point of view, it makes more sense to work with the pair groupoid $V
\times V$ because, as a groupoid, it has a richer structure.
Proposition~\ref{prop:extension} allows us to tie up both aspects by
showing that they are equivalent.  

\begin{figure}
\begin{center}
  \includegraphics[scale=0.7]{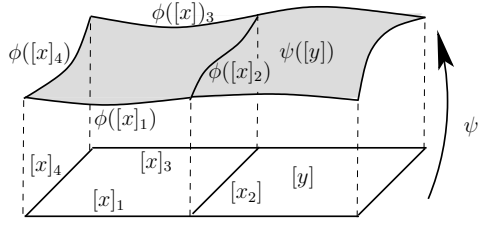}
  \caption{A discrete field $\phi$ and its associated mapping $\psi:
    \bfX^k \rightarrow \bfG^k$}
  \label{fig:section}
\end{center}
\end{figure}

It now remains to make the link between discrete fields, or morphisms
of groupoids, on the one hand, and mappings from $\bfX^k$ to $\bfG^k$
on the other hand.  It is straightforward to see that a morphism
$\phi: V \times V \rightarrow G$ induces a map $\psi: \bfX^k
\rightarrow \bfG^k$ by putting
\be \label{induce}
  \psi([x]) = ( \phi([x]_1), \ldots, \phi([x]_k) ).
\ee
(see also figure~\ref{fig:section}).  The map $\psi$ has some
properties reminiscent of those of groupoid morphisms.  Of particular
importance is the following:

\textbf{Morphism property:} if $[x]$ and $[y]$ are elements of
$\bfX^k$ having an edge in common, then the images of $[x]$ and $[y]$
under $\psi$ have the corresponding edge in $\bfG^k$ in common.  Explicitely:
\be \label{def:morphismX}
[x]_l = ([y]_m)^{-1} \quad \text{ implies that } \quad \psi([x])_l =
(\psi([y])_m)^{-1}.
\ee

\begin{prop} \label{prop:onetoone}
  There is a one-to-one correspondence between groupoid morphisms
  $\phi: V \times V \rightarrow G$ and mappings $\psi: \bfX^k
  \rightarrow \bfG^k$ satisfying the morphism property.
\end{prop}
\begin{mproof}
We have already associated with a groupoid morphism $\phi$ a map
$\psi$ satisfying the morphism property.  To prove the converse, let
$\psi: \bfX^k \rightarrow \bfG^k$ be a map satisfying the morphism
property.  Define first $\phi: E \rightarrow G$ as follows.

\begin{enumerate}
\item For $(u, u) \in E$, we take a $k$-gon $[x]$ having $u$ as its
  $l$th vertex: $u = \alpha_X( [x]_l )$ and we put
\[
  \phi(u, u) = \alpha^{(l)}( \psi([x]) ).
\]
It is straightforward but rather tedious to show that this expression
does not depend on the choice of $[x]$.  Let $[y]$ be another $k$-gon,
with $u$ as its $m$th vertex.  Let us assume for the sake of
simplicity that $u$ has degree four (the general case can be dealt
with by repeated application of this special case).  Then the edges
that emerge from $u$ are $[x]_l$ and $[y]_m$, as well as
$([x]_{l-1})^{-1}$ and $([y]_{m-1})^{-1}$ (see
figure~\ref{fig:vertex}) and there exists exactly one $k$-gon $[z]$
such that
\[
  [z]_n = ([x]_l)^{-1} \qqand [z]_{n+1} = ([y]_{m-1})^{-1}.
\]
By definition, we have that $\beta( \psi([z])_n ) = \alpha(
\psi([z])_{n+1} )$ and $\beta( \psi([y])_{m-1} ) = \alpha( \psi([y])_m
)$.  On the other hand, the morphism property ensures that
\[
  \psi([x])_l = (\psi([z])_n)^{-1} \qqand \psi([y])_{m-1} =
  (\psi([z])_{n+1})^{-1}.  
\]
By applying $\alpha$ to the left equality and $\beta$ to the right
equality, we finally obtain that
\[
  \alpha^{(l)}( \psi( [x] ) ) = \alpha^{(m)}(\psi([y])),
\]
which shows that $\phi(u, u)$ does not depend on $[x]$.

\begin{figure}
\begin{center}
  \includegraphics[scale=0.7]{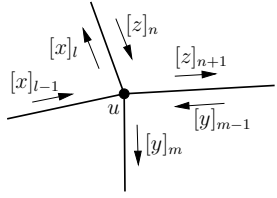}
   \caption{A vertex of degree four.}
   \label{fig:vertex}
\end{center}
\end{figure}

\item For $(u, v) \in E$, $u \ne v$, we take $[x]$ in $\bfX^k$ such
  that $(u, v) = [x]_l$ and we put
\[
  \phi(u, v) = \psi([x])_l.
\]
This is well defined because of the morphism property and, moreover,
$\phi$ satisfies $\phi(y, x) = (\phi(x, y))^{-1}$.
\end{enumerate}
By applying proposition~\ref{prop:extension} we obtain the desired
morphism $\phi: V \times V \rightarrow G$.
\end{mproof}

Note that we can still view the developments in the preceding sections
as follows.  First, the frame bundle of $X$ was discretized by
considering the set of $k$-gons $\bfX^k$.  Secondly, the jet bundle
was discretized by essentially the same procedure: as a jet in the
continuous case can be identified with a ``horizontal'' subspace, we
discretised the jet bundle of $\pi$ by approximating jets by $k$-gons
in $G$.  Finally, we introduced discrete fields as groupoid morphisms
from $V \times V$ to $G$, or, equivalently, mappings from $\bfX^k$ to
$\bfG^k$ satisfying the morphism property.  This property can be seen
as the discrete analogue of a section of $J^1\pi$ being holonomic.

\begin{remark}
  It is perhaps useful to illustrate the theory developed so far by
  applying it to groupoid mechanics.  In this case, the base space $X$
  is $\bfR$, but all of the constructions for $X = \bfR^2$ carry
  through to this case.  As a discretization of $\bfR$, we choose the
  canonical injection $i: \bfZ \hookrightarrow \bfR$.  A discrete
  field can then be identified with a bi-infinite sequence of pairwise
  composable groupoid elements $\ldots, g_{-2}, g_{-1}, g_0, g_1,
  \ldots$, which is precisely the definition of an \emph{admissible
    sequence} in \cite{GroupoidMech05, Weinstein96}.
\end{remark}

\subsection{The prolongation $\prl$} \label{sec:prl}

We recall that the discrete jet bundle $\bfG^k$ is equipped with $k$
generalized source maps defined as $\alpha^{(i)}([g]) = \alpha([g]_i)$.
By use of these maps, we define the \emph{prolongation} $\prl$ of
$\bfG^k$ through the following commutative diagram:
\[\xymatrix{
  \prl \ar[r] \ar[d] & AG \times \cdots \times AG \ar[d] \\
  \bfG^k \ar[r] & Q \times \cdots \times Q
}\]
Hence, $\prl$ consists of elements $([g]; v_1, \ldots, v_k)$, where
$v_i \in A_{\alpha(g_i)}G$ for each $i = 1, \ldots, k$.  We denote by
$\pi^{(k)}: \prl \rightarrow \bfG^k$ the projection which maps $([g];
v_1, \ldots, v_k)$ onto $[g]$.  Furthermore, there exist $k$ bundle
morphisms $(P^{(i)}, p^{(i)}): \prl \rightarrow PG$, defined as
follows.  The base space map $p^{(i)}: \bfG^k \rightarrow G$ is the
projection onto the $i$th factor, $p^{(i)}([g]) = [g]_i$, and the
total space map $P^{(i)}$ is defined as
\[
P^{(i)}([g]; v_1, \ldots, v_k) = (g_i; v_i, v_{i+1}).
\]

The definition of $\prl$ is strongly reminiscent of that of the
prolongation of a Lie groupoid over a fibration (see
section~\ref{sec:prlgroupoid}), although in general $\bfG^k$ is not a
groupoid.  The exact nature of $\prl$ is unclear at this stage, but we
will show (see theorem~\ref{thm:algebroid}) that the algebroid
structure of $PG$ can be used to equip $\prl$ with a Lie algebroid
structure by demanding that the maps $(P^{(i)}, p^{(i)})$ are
Lie-algebroid morphisms.

\begin{remark} \label{remark:diffeo} For $k = 2$, the manifold
  $\bfG^2$ is diffeomorphic to $G$, with the diffeomorphism $\varphi$
  mapping each pair $(g, g^{-1})$ onto $g$.  Note that $p^{(1)} =
  \varphi$.  In addition, we have that
  \[
  \alpha^{(1)} = \alpha \circ \varphi \qqand \alpha^{(2)} = \beta
  \circ \varphi,
  \]
  confirming our intuition that the maps $\alpha^{(i)}$ are some
  sort of ``generalized source maps''.  Furthermore, the projection
  $P^{(1)}$ is given by
  \[
  P^{(1)}(g, g^{-1}; u_{\alpha(g)}, v_{\beta(g)}) = (g; u_{\alpha(g)},
  v_{\beta(g)}),
  \]
  and so in fact it is just the natural identification of $\prl[2]$
  with $PG$.  On the other hand, $P^{(2)}$ is given by 
  \[
  P^{(2)}(g, g^{-1}; u_{\alpha(g)}, v_{\beta(g)}) = (g^{-1}; v_{\beta(g)},
  u_{\alpha(g)}).
  \]
  We recalled in section~\ref{sec:prlgroupoid} that $PG$ is a groupoid
  over $AG$ in a natural way.  A brief comparison shows that $P^{(2)}$
  is just the inversion mapping of $PG$, once we use $P^{(1)}$ to
  identify $PG$ and $\prl[2]$.  
\end{remark}

\subsubsection{The injection $\calI: \prl \hookrightarrow
  T\bfG^k$} \label{sec:inject} 

Of central importance for the following developments is the fact that
there exists a bundle injection $\calI$ of $\prl$ into $T \bfG^k$.
In order to define $\calI$, we recall that a section $v$ of the Lie
algebroid $AG$ defines on $G$ a left-invariant vector field $v^L$ and
a right-invariant vector field $v^R$ (see expression
(\ref{vectorfields})).  We also recall that we use the same notation
for the pointwise operation (see remark~\ref{remark:notation}).

Now, let $([g]; v_1, \ldots, v_k)$ be any element of $\prl$, 
and define
$\calI([g]; v_1, \ldots, v_k) \in T_{[g]} \bfG^k$ as 
\[
  \calI([g]; v_1, \ldots, v_k) = (v_1^R(g_1) + v_2^L(g_1), v_2^R(g_2)
  + v_3^L(g_2), \ldots, v_k^R(g_k) + v_1^L(g_k)).
\]
To prove that the right-hand side is a tangent vector to $\bfG^k$ at $[g]$, we
take for each $i = 1, \ldots, k$ a curve $t \mapsto h_i(t) \in
\calF^\alpha(g_i)$ in the $\alpha$-fibre through $g_i$ such that
$h_i(0) = \alpha(g_i)$ and $\dot{h}_i(0) = v_i$.  Then the vector on
the right-hand side is the tangent vector at $0$ to the following
curve in $\bfG^k$:
\[
  t \mapsto \big( h_1^{-1}(t)g_1h_2(t), h_2^{-1}(t)g_2h_3(t), \ldots,
  h_k^{-1}(t)g_kh_1(t) \big).
\]

\begin{definition} \label{def:lift}
  Let $[g]$ be an element of $\bfG^k$.  The \emph{$i$th tangent lift}
  is the map $L^{(i)}_{[g]}: A_{\alpha(g_i)}G \rightarrow
  T_{[g]}\bfG^k$ defined as 
  \[
     L^{(i)}_{[g]}(v) = \calI([g]; 0, \ldots, 0, v, 0, \ldots, 0)
     \qfor v \in A_{\alpha(g_i)}G,
  \]
  where $v$ occupies the $i$th position among the arguments of
  $\calI([g]; \ldots)$.  We will frequently use the notation
  $v^{(i)}_{[g]}$ for the element $L^{(i)}_{[g]}(v)$.
\end{definition}

\begin{remark} \label{remark:inj}
  We pointed out that $\prl[2]$ is isomorphic to $PG$.  In
  this case, the injection $\calI$ is given by 
  \[
    \calI: (g; u_{\alpha(g)}, v_{\beta(g)}) \mapsto T(r_g \circ
    i)(u_{\alpha(g)}) + Tl_g(v_{\beta(g)}) \in V_g \beta \oplus V_g
    \alpha, 
  \] 
  and coincides with the isomorphism $\Theta: PG \rightarrow V\beta
  \oplus V\alpha$ (see section~\ref{sec:prlgroupoid}).  In this case,
  the map $\calI$ can also be seen as the anchor of the Lie algebroid
  $PG$.  This theme will return in the next section, when we endow
  $\prl$ with the structure of a Lie algebroid, with $\calI$ as its
  anchor map.
\end{remark}

\subsubsection{The Lie algebroid structure on $\pi^{(k)}: \prl
  \rightarrow \bfG^k$}

In order to endow $\prl$ with the structure of a Lie algebroid, we
introduce the concept of the \emph{lift of a section} of $PG$ to
$\prl$, not to be confused with the tangent lift of
definition~\ref{def:lift} (In fact, the lift operation defined here
will be used only in this section).

We recall that a pair of sections $u, v$ of $AG$ induces a section $X$
of $PG \rightarrow G$ according to $X(g) = (g; u_{\alpha(g)},
v_{\beta(g)})$.  The Lie bracket on $PG$ is completely determined by
its action on sections of this form (see section~\ref{sec:prlPG}).  We
now define the $i$th \emph{lift of $X$} as the section $X_{(i)}$ of $\prl$
constructed as follows:
\be \label{Xlift}
  X_{(i)}([g]) = ( [g]; 0, \ldots, 0,
    \underbrace{u_{\alpha(g_i)}}_{i},
    \underbrace{v_{\alpha(g_{i+1})}}_{i+1}, 0, \ldots, 0 ) \qfor [g]
    \in \bfG^k.
\ee
We will show that $\prl$ can be equipped with the structure of a Lie
algebroid over $\bfG^k$, and that its Lie bracket is completely
determined by its action on sections $X_{(i)}$ of the form
(\ref{Xlift}). 

\begin{theorem} \label{thm:algebroid}
  There exists a unique Lie algebroid structure on $\pi^{(k)}: \prl
  \rightarrow \bfG^k$ such that each projection map $P^{(i)}: \prl
  \rightarrow PG$ is a morphism of Lie algebroids.  This Lie algebroid
  structure is characterized by 
  \begin{enumerate}
  \item the anchor $\rho^{(k)} : \prl \rightarrow T\bfG^k$ coincides
    with the injection $\calI$ defined in section~\ref{sec:inject};
  \item for $X, Y \in \Sec(AG)$ and $X_{(i)}, Y_{(i)}$ the
    corresponding $i$th lifts, the bracket of $X_{(i)}$ and $Y_{(i)}$
    is determined by \be \label{bracket} P^{(i)} \circ [X_{(i)},
    Y_{(i)}] = [X, Y] \circ p^{(i)}.  \ee
  \end{enumerate}
  We denote the associated exterior differential on
  $\bigwedge(\prl)^\ast$ by $\d^{(k)}$.   
\end{theorem}
\begin{mproof}
As all of the projection mappings $(P^{(i)}, p^{(i)})$ are Lie
algebroid morphisms, the anchor $\rho^{(k)}$ of $\prl$ satisfies
\[
  \hat{\rho} \circ P^{(i)} = Tp^{(i)} \circ \rho^{(k)},
\]
where $\hat{\rho}: PG \rightarrow TG$ is the anchor of $PG$ (see
section~\ref{sec:prlPG}).  Hence, the $i$th component of
$\rho^{(k)}([g]; v_1, \ldots, v_k)$ is just $\rho(g_i; v_i, v_{i+1})$,
which is equal to $v_i^R(g_i) + v_{i+1}^L(g_i)$.  We conclude that
$\rho^{(k)}$ is precisely the injection $\calI$.

The $i$th lift $X_{(i)}$ of $X$ satisfies 
\be \label{decomp}
  P^{(i)} \circ X_{(i)} = X \circ p^{(i)}
\ee
and the bracket of $X_{(i)}$ and $Y_{(i)}$ is therefore given by the
corresponding expression in (\ref{bracket}).  This follows from
\cite[def.~1.3]{Higgins90} by noting that (\ref{decomp}) is the
$P^{(i)}$-decomposition of $X_{(i)}$.  It is easy to see that the
bracket of two $i$th lifts is uniquely determined by (\ref{bracket}).
That the bracket of two arbitrary sections of $\pi^{(k)}$ is also
determined by this expression, is a consequence of the fact that one
may lift a basis $\{e_\alpha\}$ of sections of $PG$ to yield a basis
$\{(e_\alpha)_{(i)}\}$ of sections of $\pi^{(k)}$.
\end{mproof}

\section{Lagrangian field theories}

After the discussion in the previous sections of the geometrical
background for our treatment of discrete field theories, we now turn
to the fields themselves, as well as the equations that govern their
behaviour.  These equations will turn out to be (implicit or explicit)
difference equations.

The key element in constructing these discrete field equations is the
specification of a \emph{discrete Lagrangian}, i.e. a smooth function $L$
on $\bfG^k$.  Associated to such a discrete Lagrangian 
is an action sum --- the
discrete counterpart of the action integral in continuous field
theory.  As we will see, the discrete field equations arise by
extremizing (in some suitable sense) this action sum.

Before deriving the discrete field equations, we will first construct
some intrinsic objects on the prolongation bundle $\pi^{(k)}: \prl
\rightarrow \bfG^k$ and we will argue that all of these objects have a natural
counterpart in continuous field theories.  These include, among other
things, the Poincar\'e-Cartan forms and the induced Legendre
transformations.  In \S~\ref{sec:examples}, we will make the link with
\cite{MPS98} when we turn our attention to an important special case:
that of the pair groupoid $G = Q \times Q$.

\subsection{The Poincar\'e-Cartan forms} \label{sec:PC}

Let $L: \bfG^k \rightarrow \bfR$ be a discrete Lagrangian.  To $L$ one
can associate $k$ sections $\theta^{(i)}_L$ of $(\pi^{(k)})^\ast :
(\prl)^\ast \rightarrow \bfG^k$, called \emph{Poincar\'e-Cartan
  forms}, which are defined as follows:
\[
  \theta^{(i)}_L([g]; v_1, \ldots, v_k) = (v_i^{(i)})_{[g]}(L), 
\]
where $v_i \in A_{\alpha(g_i)}G$ and 
$v^{(i)}_i$ is the $i$th tangent lift of $v_i$ to $\bfG^k$
(cf. definition~\ref{def:lift}).  As
$\sum v^{(i)}_i = \calI([g]; v_1, \ldots, v_k)$, we may conclude that
\[
  \d^{(k)} L = \sum_{i = 1}^k \theta^{(i)}_L.
\]

\begin{remark}
  In the case $k = 2$, it follows from remark~\ref{remark:inj} that
  $\theta^{(1)}_L$, resp. $\theta^{(2)}_L$, can be identified with the
  Poincar\'e-Cartan forms $\theta^-_L$, resp. $\theta^+_L$, defined in
  \cite{GroupoidMech05} as
  \[
    \theta^-_L(g; u_{\alpha(g)}, v_{\beta(g)}) = \d L(g)(u^R(g)) \qqand
    \theta^+_L(g; u_{\alpha(g)}, v_{\beta(g)}) = \d L(g)(v^L(g)).
  \]

  Indeed, let us consider the function $L_{\mathrm{mech}}$ on $G$
  given by $L_{\mathrm{mech}} = \varphi_\ast L$, where $\varphi:
  \bfG^2 \rightarrow G$ is the diffeomorphism introduced in
  remark~\ref{remark:diffeo}, or, explicitely, $L_{\mathrm{mech}}(g) =
  L(g, g^{-1})$.  Then, by definition,
  \[
    \theta_L^{(1)}(g, g^{-1}; u_{\alpha(g)}, v_{\beta(g)}) = \ddt
    L(h^{-1}(t)g, g^{-1}h(t)) \Big|_0,
  \]
  where $h(t) \in \calF^\alpha(g)$ is such that $h(0) = \alpha(g)$ and
  $\dot{h}(0) = u_{\alpha(g)}$.  The right-hand side can now be
  rewritten as 
  \[
  \ddt L_{\mathrm{mech}}(h^{-1}(t)g) \Big|_0 = \left< \d
    L_{\mathrm{mech}}, T(r_g \circ i)(u_{\alpha(g)}) \right> =
  \theta^-_L(g; u_{\alpha(g)}, v_{\beta(g)}).
  \]
  There is a similar identification of $\theta^{(2)}_L$ with
  $\theta^+_{L_{\mathrm{mech}}}$. 
\end{remark}

\subsection{The field equations} \label{sec:FE}

We now proceed to derive the discrete field equations for a Lie
groupoid morphism
$\phi: V \times V \rightarrow G$ by varying a discrete action sum.  Let
$L: \bfG^k \rightarrow \bfR$ be a discrete Lagrangian and define the
\emph{action sum} as 
\be \label{action}
   S(\phi) = \sum_{[x] \in \bfX^k} L( \psi([x]) ),
\ee
where $\psi$ is the map from $\bfX^k$ to $\bfG^k$ associated to the
morphism $\phi$ (see proposition~\ref{prop:onetoone}).  Strictly
speaking, one should take care to ensure that this summation is finite
by restricting to morphisms $\phi$ whose domain of definition $U
\subset V \times V$ only contains a finite number of edges.

\subsubsection{Variations}

In this section, we define the concept of a \emph{variation}, both
finite and infinitesimal.  A key property is that the variation of a
groupoid morphism yields a new groupoid morphism.  In order to
formalize this, we introduce the concept of morphism properties for
mappings from $\bfG^k$ onto itself.  These properties are very similar
to the morphism property introduced in (\ref{def:morphismX}).

Let us introduce a slight modification of the source mappings
$\alpha^{(i)}$:
\[
  \hat{\alpha}^{(i)}: \bfG^k \rightarrow \bfG^k, \quad
  \hat{\alpha}^{(i)}([g]) = (\alpha([g]_i), \ldots, \alpha([g]_i)). 
\]
It is obvious that for any $l \le k$, $\left( \hat{\alpha}^{(i)}([g])
\right)_l = \alpha^{(i)}([g])$.

\begin{definition}
  A map $\Psi: \bfG^k \rightarrow \bfG^k$ is said to satisfy the
  \emph{morphism properties} if, for all $[g], [h] \in \bfG^k$,
  \begin{RomanList}
    \item $\Psi \circ \hat{\alpha}^{(i)} = \hat{\alpha}^{(i)} \circ
      \Psi$ for $i = 1, \ldots, k$; \label{MPunit}
    \item if $[g]_l = [h]_m$, then $\Psi([g])_l =
      \Psi([h])_m$. \label{MPedge} 
  \end{RomanList}
\end{definition}

\begin{prop}
  There is a one-to-one correspondence between groupoid morphisms
  $\Phi: G \rightarrow G$ and mappings $\Psi: \bfG^k \rightarrow
  \bfG^k$ satisfying the morphism properties. 
\end{prop}
\begin{mproof}
Let $\Phi$ be a morphism from $G$ to itself.  As in (\ref{induce}),
$\Phi$ induces a mapping $\Psi: \bfG^k \rightarrow \bfG^k$ satisfying
the morphism properties, namely:
\[
  \Psi([g]) = \left( \Phi([g]_1), \ldots, \Phi([g]_k) \right).
\]

Conversely, let $\Psi: \bfG^k \rightarrow \bfG^k$ be a mapping
satisfying the morphism properties and let $g$ be any element of $G$.
In order to define $\Phi(g)$, we take any $[\eta] \in \bfG^k$ such that
there exists a natural number $l \le k$ for which $g = [\eta]_l$.  We
then put 
\[
    \Phi(g) := \Psi([\eta])_l.
\]
Morphism property \ref{MPedge} ensures that $\Phi(g)$ depends only on
$g$ and not on the other components of $[\eta]$.  We now have to
check that $\Phi$ is a morphism of $G$ to itself.

\begin{enumerate}
\item In order to prove that $\alpha \circ \Phi = \Phi \circ \alpha$,
  we take any $g \in G$ and consider $[\eta] \in \bfG^k$ such that $[\eta]_l
  = g$.  Then $\alpha(\Phi(g)) = \alpha(\Psi([\eta])_l) =
  \alpha^{(l)}(\Psi([\eta]))$. 

  However, because of morphism property \ref{MPunit} we have
  \be \label{alphacommute}
    \hat{\alpha}^{(l)}\big(\Psi([\eta])\big) 
    = \Psi\big(\hat{\alpha}^{(l)}([\eta])\big)
    = \Psi\big( (\alpha(g), \ldots, \alpha(g)) \big).
  \ee
  For any arbitrary $m \le k$, we have that $\Phi( \alpha(g) ) =
  \Psi\big((\alpha(g), \ldots, \alpha(g))\big)_m$, and so, by
  considering the $m$th component of (\ref{alphacommute}), 
  \begin{align*}
    \Phi(\alpha(g)) & = \left( \hat{\alpha}^{(l)}\big( \Psi([\eta]) \big)
    \right)_m \\
    & = \alpha^{(l)}\big( \Psi([\eta]) \big),
  \end{align*}
  from which we conclude that $\alpha(\Phi(g)) = \Phi(\alpha(g))$
  for all $g \in G$.  A similar argument can be used to show that
  $\Phi$ commutes with $\beta$.
 
\item We now show that $\Phi(g^{-1}) = \Phi(g)^{-1}$ for any $g \in
  G$.  Let 
  \[
    [\xi] = (g, g^{-1}, \alpha(g), \ldots, \alpha(g)), 
  \]
  then $\Psi([\xi])_1 = \Phi(g)$, $\Psi([\xi])_2 = \Phi(g^{-1})$ and
  $\Psi([\xi])_j = \Phi(\alpha(g))$ for $j = 3, \ldots, k$.  Moreover,
  since $\Psi([\xi]) \in \bfG^k$, we have, by definition of $\bfG^k$,
  that $\Psi([\xi])_1 \cdots \Psi([\xi])_k = \alpha(\Psi([\xi])_1)$,
  or
  \[
    \Phi(g) \Phi(g^{-1}) \Phi(\alpha(g)) \cdots \Phi(\alpha(g)) =
    \alpha( \Phi(g) ),
  \]
  which, after simplication, leads to $\Phi(g^{-1}) = \Phi(g)^{-1}$.

\item Finally, we have to show that if $(g, h)$ is a composable pair,
  i.e. $\beta(g) = \alpha(h)$, then $(\Phi(g), \Phi(h))$ is also
  composable, and moreover, $\Phi(gh) = \Phi(g) \Phi(h)$.  The proof
  of this property is similar to the proof of the previous property.

  Consider the following $k$-gon:
  \[
    [\eta] = (g, h, (gh)^{-1}, \alpha(g), \ldots, \alpha(g)).
  \]
  Then, as $\Psi([\eta]) \in \bfG^k$, we conclude that, first of all,
  $\beta(\Phi(g)) = \alpha(\Phi(h))$, and secondly 
  \[
    \Phi(g) \Phi(h) \Phi((gh)^{-1}) = \alpha\big( \Phi(g) \big).
  \]
  By using the previous properties, as well as some of the standard
  properties of the groupoid $G$, we find that $\Phi(gh) = \Phi(g)
  \Phi(h)$.
\end{enumerate} 
We conclude that $\Phi: G \rightarrow G$ is a groupoid morphism.
\end{mproof}

\begin{corollary}
  Let $\Psi : \bfG^k \rightarrow \bfG^k$ be a map satisfying the
  morphism properties.  Then for each $[g] \in \bfG^k$,
  \[
    \Psi([g]^{-1}) = \Psi([g])^{-1}.
  \]
\end{corollary}
\begin{mproof}
This can be proved directly, or by noting that $\Psi$ induces a
groupoid morphism $\Phi$ such that
\[
  \Psi([g]) = \big(\Phi([g]_1), \ldots, \Phi([g]_k)\big),
\]
and writing out the definition of $[g]^{-1}$ and $\Psi([g])^{-1}$.
\end{mproof}

After these introductory lemmas, we now turn to the concepts of finite
and infinitesimal variations of a morphism $\phi: V \times V
\rightarrow G$.  Before doing so, we remark that any subset $\hat{U}$
of $\bfX^k$ uniquely determines a subset $U$ of $V \times V$,
consisting of all the edges of all faces contained in $\hat{U}$. We
then define the \emph{boundary} $\partial U \subset V \times V$ to be
the following set:
\begin{gather*}
  \partial U := \big\{ (u, v) \in V \times V : \text{$\exists [x], [y]
    \in \bfX^k$ such that $[x]_l = (u, v), [y]_m = (v, u)$} \\
    \text{and $[x] \in \hat{U}, [y] \notin \hat{U}$} \big\}.
\end{gather*}
In other words, the boundary $\partial U$ consists of edges that, when
traversed in opposite directions, are part of two $k$-gons $[x]$ and
$[y]$, one of which is contained in $\hat{U}$, while the other one is
not.

\begin{definition} \label{def:finvar} A \emph{finite variation} over
  $\hat{U} \subset \bfX^k$ of a morphism $\phi: V \times V \rightarrow
  G$, with associated mapping $\psi: \bfX^k \rightarrow \bfG^k$, is a
  map $\Psi: \bfR \times \psi(\hat{U}) \rightarrow \psi(\hat{U})$ such
  that for each fixed $t \in \bfR$, $\Psi_t := \Psi(t, \cdot)$
  satisfies the morphism properties, and which has the following form:
  for each $[g] \in \psi(\hat{U})$ there exist maps $h_i: \bfR
  \rightarrow G$ such that
  \be \label{varg}
     \Psi_t([g]) := \Psi(t, [g]) = (h_1(t)^{-1}g_1h_2(t), h_2(t)^{-1}g_2h_3(t),
     \ldots, h_k(t)^{-1}g_kh_1(t)),
  \ee
  where $h_i(0) = \alpha(g_i)$ and $h_i(t) \in \calF^\alpha(g_i)$.  In
  addition, if $[g]_l \in \phi(\partial U)$, then $h_l(t) =
  \alpha([g]_l)$ and $h_{l+1}(t) = \beta([g]_l)$ for all $t \in \bfR$.
\end{definition}

Note that $\Psi_t$ doesn't have to be defined on the whole on
$\bfG^k$, but only on the image of $\hat{U}$ under $\psi$.  Note
furthermore that $\Psi_0$ is the identity mapping on $\psi(\hat{U})$,
since each of the curves $h_i: \bfR \rightarrow G$ in (\ref{varg})
satisfies $h_i(0) = \alpha(g_i)$.

\begin{remark}  
  It should be emphasised that in (\ref{varg}), each of
  the curves $h_i: \bfR \rightarrow \calF^\alpha(g_i)$ depends only on
  $\alpha(g_i)$, and not on the whole of $[g]$ as might be expected.
  In order to prove this, consider the morphism $\Phi_t$ associated to
  the variation $\Psi_t$ and let $g, g'$ be elements of $G$ such that
  $\beta(g) = \alpha(g')$.  Then
  \[
    \Phi_t(g) = h(t)^{-1}gk(t) \qqand \Phi_t(g') = h'(t)^{-1}g'k'(t),
  \]
  which makes it clear that $h(t), k(t)$ can only depend on $g$, and
  $h'(t), k'(t)$ only on $g'$.  However, as $(g, g')$ is a composable
  pair, so is their image under $\Phi_t$ and therefore $h'(t) = k(t)$.
  We conclude that $k(t)$ cannot depend on $g$ itself but only on
  $\alpha(g)$.  For the variation (\ref{varg}), a similar argument
  implies that each $h_i$ only depends on $\alpha(g_i)$.
\end{remark}

\begin{remark} \label{remark:composable} Let $[x], [y] \in \bfX^k$ be
  two $k$-gons that have an edge in common, e.g. $[x]_l = [y]_m^{-1}$
  for $l, m \le k$.  Consider now their images under the mapping
  $\psi$ associated to a morphism $\phi: V \times V \rightarrow G$,
  namely $[\eta] = \phi([x])$ and $[\xi] = \phi([y])$.  Because of
  the morphism property, we conclude that $[\eta]_l = [\xi]_m^{-1}$.
  Moreover, $([\eta]_{l-1}, [\xi]_{m+1})$ is a composable pair, as is
  $([\xi]_{m-1}, [\eta]_{l+1})$.  Let $\Psi$ be a variation of $\phi$; it is
  interesting to compare its action on $[\eta]$ and $[\xi]$.
  Putting
  \[
  \Psi_t([\eta]) = (\ldots, h_{l-1}(t)^{-1} \eta_{l-1} h_l(t),
  h_l(t)^{-1}\eta_lh_{l+1}(t), h_{l+1}(t)^{-1}\eta_{l+1}h_{l+2}(t),
  \ldots),
  \]
  where we denote $[\eta]_i$ simply by $\eta_i$, $i = 1, \ldots, k$,
  the morphism properties that $\Psi_t$ has to satisfy, allow us to
  conclude that the variation of $[\xi]$ is given by the form
  \[
  \Psi_t([\xi]) = (\ldots, k_{m-1}(t)^{-1}\xi_{m-1}h_{l+1}(t),
  h_{l+1}(t)^{-1}\xi_m h_l(t), h_l(t)^{-1}\xi_{m+1}k_{m+2}(t), \ldots).
  \]
  The important thing to note is that a composable pair, for example
  $([\eta]_{l-1}, [\xi]_{m+1})$, is mapped to another composable pair,
  in this case $(\Psi_t([\eta])_{l-1}, \Psi_t([\xi])_{m+1})$.
\end{remark}

\begin{figure}
\begin{center}
  \includegraphics[scale=0.8]{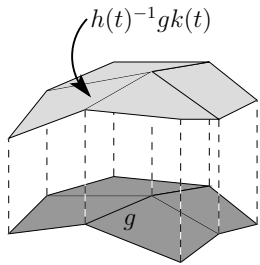}
  \caption{A variation of a discrete field (boundary not shown).}
  \label{fig:var}
\end{center}
\end{figure}

In conclusion, although definition~\ref{def:finvar} might seem quite
involved at first, it has nevertheless a clear geometric
interpretation.  Indeed, each edge $g$ in the image of $U$ under the
discrete field $\phi$ is varied according to the following
prescription: there exist curves $h(t)$ and $k(t)$ in
$\calF^\alpha(g)$, with $h(0) = k(0) = \alpha(g)$ such that the
variation of $g$ can be expressed as
\[
  g \mapsto h(t)^{-1}gk(t).
\]
The edges of the boundary $\partial U$ are not varied.  Imposing the
morphism condition on $\Psi_t$ ensures us that each edge is varied in
a uniquely determined way, and moreover, composable edges (i.e. having
a vertex in common) are mapped to composable edges.  We have sketched
the effect of a finite variation on a discrete field in
figure~\ref{fig:var}.

\begin{definition} \label{def:infvar}
  An \emph{infinitesimal variation} over $\hat{U} \subset \bfX^k$ of a
  morphism $\phi: V \times V \rightarrow G$ is a
  section $\Gamma$ of $\pi^{(k)}$, defined on $\psi(\hat{U})$, such that 
  \begin{enumerate}
    \item $[g]_l = [h]_m$ implies that $\Gamma([g])_l =
      \Gamma([h])_m$; 
    \item $[g]_l \in \phi(\partial U)$ implies that $\Gamma([g])_l = 0$, 
  \end{enumerate}
  (with the convention that for $\Gamma([g]) = ([g]; v_1, \ldots,
  v_k)$, $\Gamma([g])_l = v_l$).
\end{definition}

In this definition, the first property ensures that $\Gamma$
attributes a unique Lie algebroid element to each edge, whereas the
second property expresses the fact that $\Gamma$ is zero on the image
of $\partial U$ under $\phi$.  We may therefore conclude that, because
of the additional conditions in definition~\ref{def:infvar}, an
infinitesimal variation can also be interpreted as a section
of $PG \rightarrow G$ defined on $\phi(U)$, or equivalently, a section
of $\phi^\ast PG$ which is zero on $\partial U$.

The infinitesimal variation $\Gamma$ associated to a finite variation
$\Psi_t$ is generated as follows.  For $[g] \in \psi(\hat{U})$,
consider the curves $h_i: \bfR \rightarrow G$
(cf. definition~\ref{def:finvar}) and put
\[
  \Gamma([g]) = ([g]; v_1, \ldots, v_k), \where v_i = \dot{h}_i(0).
\]
The infinitesimal variation $\Gamma$ will satisfy the required
conditions since $\Phi_t$ has the morphism properties and leaves 
the image of $\partial U$ invariant.

Conversely, we may ``integrate'' an infinitesimal variation $\Gamma$
to yield a finite variation.  Let $\gamma$ be the section of $PG$
associated to $\Gamma$, which can be written as $\gamma(g) = (g;
u_{\alpha(g)}, v_{\beta(g)})$, where $u_{\alpha(g)} \in
A_{\alpha(g)}G$ and $v_{\beta(g)} \in A_{\beta(g)}G$.  Consider now
the left- and right-invariant vector fields $u^L$ and $v^R$, defined
as $u^L(g) = (u_{\alpha(g)})^L(g)$, and $v^R = (v_{\beta(g)})^R(g)$.
Let $\theta: \bfR \times G \rightarrow G$ the flow of $u^L$, and
$\varphi$ the flow of $v^R$.  We then define a morphism $\Phi_t: G
\rightarrow G$ by putting $\Phi_t(g) = \varphi_t(g)g\theta_t(g)$.  It
is easy to check that this composition is well defined.  Now, the
morphism $\Phi_t$ induces a finite variation in the sense of
definition~\ref{def:finvar} and by construction the associated
infinitesimal variation is equal to the original section $\Gamma$.

\subsubsection{The field equations}

For the sake of clarity, we will derive the field equations in the
case where $X (= \bfR^2)$ is covered by a quadrangular mesh as in
figure~\ref{fig:quadrangles}.  This is also one of the cases covered
in \cite{MPS98}.  The generalization of the field equations to
non-regular meshes is straightforward but involves a lot of notational
intricacies.

Let $\phi: V \times V \rightarrow G$ be a discrete field, with
associated mapping $\psi: \bfX^k \rightarrow \bfG^k$.  Consider a
finite subset $\hat{U}$ of $\bfX^k$ with its induced set $U \subset V
\times V$, and let $\Psi_t$ be a finite
variation (according to definition~\ref{def:finvar}) over $\hat{U}$ of
$\phi$.  We denote the composition $\Psi_t \circ \psi$ as $\psi_t$.
Because $\Psi_t$ satisfies the morphism properties, $\psi_t$ induces
in turn a groupoid morphism $\phi_t : V \times V \rightarrow G$.  Note
that $\phi_0 = \phi$.

We now express that the (restricted) morphism $\phi: (V \times V) \cap
U \rightarrow G$ extremizes the action sum (\ref{action}), i.e.
\be \label{actionsum}
  \ddt S(\phi_t)\Big|_{t = 0} = 0
\ee
for an arbitrary variation $\Psi_t$.

\begin{figure}
\begin{center}
  \includegraphics[scale=0.9]{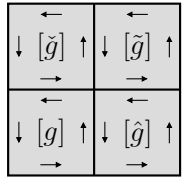}
  \caption{Schematic representation of $[g]$, $[\hat{g}]$,
    $[\tilde{g}]$, and $[\check{g}]$.}
  \label{fig:variation}
\end{center}
\end{figure}

Let $u$ be a vertex in $V$; naturally, $u$ is a common vertex of four
quadrangles, denoted by $[x]$, $[\hat{x}]$, $[\tilde{x}]$ and
$[\check{x}]$.  Let us denote by $[g]$, $[\hat{g}]$, $[\tilde{g}]$ and
$[\check{g}]$ their corresponding images under $\psi$ (see
figure~\ref{fig:variation} for a schematic representation of these
four quadrangles).  We will focus on the variation of the image of the
center vertex $u$.

The variation $\Psi_t$ will map $[g]$ into a new quadrangle
$[g']$ of the following form:
\be \label{variation}
[g'] = (h_1^{-1}(t)g_1h_2(t), h_2^{-1}(t)g_2h_3(t),
h^{-1}_3(t)g_3h_4(t), h_4^{-1}(t) g_4 h_1(t)),
\ee
and likewise for $[\hat{g}]$, $[\tilde{g}]$ and $[\check{g}]$.
However, the latter three each have a vertex in common with $[g]$, and
because of morphism property \ref{MPedge}, their variations will be
related, as we pointed out in remark~\ref{remark:composable}.  More
precisely, let us focus on the effects of $h_3(t)$: the terms in the
action sum involving $h_3(t)$ are spelled out below:
\begin{align*}
  S(\phi_t) = \cdots & + L(h_1^{-1}(t)g_1h_2(t), h_2^{-1}(t)g_2h_3(t),
   h_3^{-1}(t)g_3h_4(t), h_4^{-1}(t) g_4 h_1(t)) \\
     & + L(h_2(t) \hat{g}_1 k_1(t), k_1^{-1}(t)\hat{g}_2k_3(t),
     k_3^{-1}(t)\hat{g}_3 h_3(t), h_3^{-1}(t)g_4h_2(t)) \\
     & + L(h_3^{-1}(t)\tilde{g}_1 k_3(t), k_3^{-1}(t)\tilde{g}_2
     l_3(t), l_3^{-1}(t)\tilde{g}_3l_4(t), l_4^{-1}(t)\tilde{g}_4h_3(t))\\
     & + L(h_4^{-1}(t)\check{g}_1h_3(t), h_3^{-1}(t)\check{g}_2l_4(t),
     l_4^{-1}(t)\check{g}_3m_4(t), m_4^{-1}(t)\check{g}_4h_4(t)),
\end{align*}
where $h_1(t), h_2(t), h_3(t), h_4(t)$ as well as $k_2(t), k_3(t)$,
$l_3(t), l_4(t)$ and $m_4(t)$ are determined as in
def.~\ref{def:finvar}.  

It is helpful to keep in mind the following relations:
\[
  g_2 = \hat{g}_4^{-1}, \quad \hat{g}_3 = \tilde{g}_1^{-1} \quad
  \tilde{g}_4 = \check{g}_2^{-1} \qqand \check{g}_1 = g_3^{-1},
\]
expressing the fact that each of the four $k$-gons $[g], [\hat{g}],
[\tilde{g}], [\check{g}]$ has an edge in common with two of the other
$k$-gons. 

By demanding that $S$ be stationary, we obtain
\[
  v^{(3)}_{[g]}(L) + v^{(4)}_{[\hat{g}]}(L) + v^{(1)}_{[\tilde{g}]}(L)
  + v^{(2)}_{[\check{g}]}(L) = 0, 
\]
where $v \in AG$ is given by $v = \dot{h}_3(0)$, and the superscript
$i$ denotes the $i$th tangent lift of an element of $AG$ to $T\bfG^k$
(see definition~\ref{def:lift}).

In conclusion, we have the following characterization of extremals of
the action sum (\ref{action}).
\begin{theorem} \label{th:EL}
  Let $\phi: V \times V \rightarrow G$ be a groupoid morphism.  For
  any $u \in V$, consider the vertex $\alpha(g) = \phi(u, u)$ and let
  $[g]$, $[\hat{g}]$, $[\tilde{g}]$ and $[\check{g}]$ be the four
  quadrangles having the vertex $\alpha(g)$ in common (as in
  figure~\ref{fig:variation}).  

  Then $\phi$ is an extremum of the action sum (\ref{action}) if and
  only if, for each such vertex $\alpha(g)$ with associated
  quadrangles $[g]$, $[\hat{g}]$, $[\tilde{g}]$ and $[\check{g}]$, and
  for each $v \in A_{\alpha(g)} G$, the following holds:
  \be \label{FE}
  v^{(1)}_{[\tilde{g}]}(L) + v^{(2)}_{[\check{g}]}(L) +
  v^{(3)}_{[g]}(L) + v^{(4)}_{[\hat{g}]}(L) = 0.
  \ee
\end{theorem}

We refer to the expressions in (\ref{FE}) as the \emph{discrete field
  equations}.  In the case where $G$ is the pair groupoid, these
equations become (implicit or explicit) difference equations (see
\cite{MPS98}).  We will return to this case in section~\ref{sec:pair}.

\subsection{The Legendre transformation}

In this section, we introduce a notion of Legendre transformation and
use it to show that the pullback of the canonical section of a
suitable dual bundle yields the Poincar\'e-Cartan forms constructed in
section~\ref{sec:PC}.  More precisely, the Legendre transformation
will be a collection of $k$ bundle maps from $\prl$ to the bundle
$P^{\tau^\ast}(AG) \rightarrow A^\ast G$.  As sketched in
section~\ref{sec:prltangent}, the dual of the latter is equipped with
a canonical section $\theta$ and the pullback of this section by each
of the bundle maps corresponding to the Legendre transformation, will
provide the full set of Poincar\'e-Cartan forms.

We first introduce the pullback bundles $P^{(i)}(AG)$,
$i = 1, \ldots, k$, constructed by means of the following commutative
diagram:
\[\xymatrix{
  P^{(i)}(AG) \ar[d] \ar[r] & T\bfG^k \ar[d]^{T\alpha^{(i)}} \\
  AG \ar[r]_{\rho}                & TQ
}\]
The bundles $P^{(i)}(AG)$ bear the same relation to $\bfG^k$ as
$P^{\alpha}(AG)$ and $P^\beta(AG)$ to $G$.

\subsubsection{The mappings $\LAproj^{(i)}: \prl \rightarrow P^{(i)}(AG)$}

For each $i = 1, \ldots, k$, there is a natural injection
$\varphi^{(i)}: G \rightarrow \bfG^k$ defined as 
\[
  \varphi^{(i)}(g) = (\alpha(g), \ldots, \alpha(g), g, g^{-1},
  \alpha(g), \ldots, \alpha(g))
\]
where $g$ and $g^{-1}$ occupy the $i$th and the $(i+1)$th position,
respectively. 

The projections $P^{(i)}: \prl \rightarrow PG$, as
defined in section~\ref{sec:prl}, can be used to define 
projection mappings $\LAproj^{(i)}: \prl \rightarrow P^{(i)}(AG)$
by means of the composition
\[
  \LAproj^{(i)}: \prl \stackrel{P^{(i)}}{\longrightarrow} PG
  \stackrel{A(\Phi^\alpha)}{\longrightarrow} P^\alpha(AG) \stackrel{\id
    \times T\varphi^{(i)}}{\longrightarrow} P^{(i)}(AG),
\]
where $A(\Phi^\alpha): PG \rightarrow P^\alpha(AG)$ was defined in
section~\ref{sec:prlalphabeta}.

\begin{remark}
  For $k = 2$, we now show that the projections $\LAproj^{(1)}$ and
  $\LAproj^{(2)}$ can be identified with $A(\Phi^\alpha)$ and
  $A(\Phi^\beta)$, respectively.  We recall that $\prl[2]$ is
  isomorphic to $PG$ and that there is a diffeomorphism $\varphi:
  \bfG^2 \rightarrow G$ sending each $(g, g^{-1})$ to $g$ (see
  remark~\ref{remark:diffeo}).  Hence, $\varphi^{(1)}$ is just
  $\varphi^{-1}$ and $\varphi^{(2)}$ equals $\varphi^{-1} \circ i$.

  There is a natural identification of $P^{(1)}(AG)$ with
  $P^\alpha(AG)$, and of $P^{(2)}(AG)$ with $P^\beta(AG)$.  Using
  these identifications, it is straightforward to see that
  $\LAproj^{(1)}$ can be identified with $A(\Phi^\alpha)$.  The
  identification of $\LAproj^{(2)}$ with $A(\Phi^\beta)$ takes some
  more work.  Consider first the composition
  \[
    P^\alpha(AG) \stackrel{\id \times T\varphi^{(2)}}{\longrightarrow}
    P^{(2)}(AG) \cong P^\beta(AG),
  \]
  which is easily seen to be equal to $\id \times Ti$.  We then obtain
  the following for the map $\LAproj^{(2)}$, considered as a map into
  $P^\beta(AG)$:
  \begin{align*}
    \left( (\id \times Ti) \circ A(\Phi^\alpha) \circ P^{(2)} \right)&(g,
    g^{-1}; u_{\alpha(g)}, v_{\beta(g)}) \\ & = 
      (\id \times Ti \circ A(\Phi^\alpha))(g^{-1}; v_{\beta(g)},
      u_{\alpha(g)}) \\
      & = (\id \times Ti)(v_{\beta(g)}, T(r_{g^{-1}} \circ
      i)(v_{\beta(g)}) + Tl_{g^{-1}}(u_{\alpha(g)})) \\
      & = (v_{\beta(g)}, T(r_g \circ i)(u_{\alpha(g)}) +
      Tl_g(v_{\beta(g)})) \\
      & = A(\Phi^\beta)(g; u_{\alpha(g)}, v_{\beta(g)}),
  \end{align*}
  where we again refer to section~\ref{sec:prlalphabeta} for the
  definition of $A(\Phi^\beta)$.
\end{remark}

\subsubsection{Definition of the Legendre transformations}

Given a Lagrangian $L: \bfG^k \rightarrow \bfR$, there are $k$
distinguished bundle maps $(P\leg^{(i)}, \leg^{(i)})$ from $\prl$ to
the bundle $P^{\tau^\ast}(AG) \rightarrow A^\ast G$, which we will
call \emph{Legendre transformations}.

For each $i = 1, \ldots, k$, the base map $\leg^{(i)}: \bfG^k
\rightarrow A^\ast G$ is defined as follows.  For each $[g] \in
\bfG^k$, $\leg^{(i)}([g])$ is the element of $A^\ast_{\alpha(g_i)}G$
defined by
\[
  \leg^{(i)}([g])(v_{\alpha(g_i)}) = v^{(i)}_{\alpha(g_i)}(L) \qforall
  v_{\alpha(g_i)} \in A_{\alpha(g_i)} G.
\]
Recall that $v^{(i)}_{\alpha(g_i)}$ is the $i$th tangent lift of
$v_{\alpha(g_i)}$ to $T_{[g]}\bfG^k$.  The total space map $P
\leg^{(i)}: \prl \rightarrow P^{\tau^\ast}(AG)$ is defined as the
composition $(\id \times T \leg^{(i)}) \times \LAproj^{(i)}$.

\begin{prop}
  Let $\theta$ be the canonical section of $[P^{\tau^\ast}(AG)]^\ast
  \rightarrow A^\ast G$ defined in section~\ref{sec:prltangent}.
  Then, for $i = 1, \ldots, k$,
  \[
    (P\leg^{(i)}, \leg^{(i)})^\star \theta = \theta^{(i)}_L.
  \]
\end{prop}
\begin{mproof}
Let $([g]; v_1, \ldots, v_k)$ be an element of $\prl$ and consider
\be \label{legtrafoeq}
    [(P\leg^{(i)}, \leg^{(i)})^\star \theta]_{[g]}([g]; v_1, \ldots,
    v_k) = \theta_{\leg^{(i)}([g])}(P\leg^{(i)}([g]; v_1, \ldots, v_k)).
\ee
Now, the canonical section $\theta$ is defined by the following rule:
for $\alpha \in A^\ast G$ and $(v, X_\alpha)$ in
$(P^{\tau^\ast}(AG))_\alpha$, we have that $\theta_\alpha(v, X_\alpha)
= \alpha(v)$.  Noting that 
\[
  P\leg^{(i)}([g]; v_1, \ldots, v_k) = (v_i, \cdot)
\]
(the precise form of the second argument doesn't matter), the
right-hand side of (\ref{legtrafoeq}) then becomes
\[
  \leg^{(i)}([g])(v_i) = \theta^{(i)}_L([g]; v_1, \ldots, v_k),
\]
where the last equality follows by comparing the definition of the
$i$th Poincar\'e-Cartan form with the $i$th Legendre transformation. 
\end{mproof}

\subsection{Variational interpretation of the Poincar\'e-Cartan forms} 

In this section, we closely follow some of the ideas set out by
Marsden, Patrick, and Shkoller in \cite{MPS98}.  In that paper, the
authors gave a variational definition of discrete multisymplectic
field theories.  As we pointed out before, the class of field theories
that they consider corresponds to the case where the groupoid $G$
over the standard fibre $Q$ is the pair groupoid $Q \times Q$ (see
section~\ref{sec:pair}).

We now intend to redo their analysis to prove that the
Poincar\'e-Cartan forms that we defined in section~\ref{sec:PC}, also
arise when considering variations of a morphism $\phi$ over a set
$\hat{U}$ such that the boundary $\partial U$ is not fixed.  Moreover,
we use these observations to derive a criterion of multisymplecticity,
and show that the field equations (see theorem~\ref{th:EL}) are
multisymplectic in that sense.  Again, this is just an extension to
the case of an arbitrary groupoid $G$ of the definitions in
\cite{MPS98}.

\subsubsection{Arbitrary variations}

Consider a finite subset $\hat{U}$ of $\bfX^k$ with associated
boundary $\partial U$.  Now, let $\phi : V \times V$ be a morphism and
consider a finite variation $\Psi: \bfR \times \psi(\hat{U})
\rightarrow \psi(\hat{U})$ of $\phi$ over $\hat{U}$ as in
definition~\ref{def:finvar}.  However, we now also allow
\emph{nontrivial} variations of the field on the boundary $\partial
U$.

When extremizing the action sum (\ref{action}), there is now a
contribution from the interior of $U$, as well as a contribution from
the boundary $\partial U$, which takes the following form (with the
notations of section~\ref{sec:FE}): 
\[
  \ddt S(\phi_t) \Big|_{t = 0} = \sum_{[x] \cap \partial U \ne
    \varnothing} \left( \sum_{l; [x]_l \in \partial U} \Big(
    \theta^{(l)}_L( \psi([x]) ) \cdot \Gamma_{\psi([x])} \Big)
  \right), 
\]
where $\Gamma$ is the infinitesimal variation associated to $\Psi$.
Once again, we see how the Poincar\'e-Cartan forms arise naturally in
the context of discrete Lagrangian field theories.

\subsubsection{Multisymplecticity}

By exactly the same reasoning as in \cite{MPS98}, we obtain a concise
criterion for multisymplecticity.  We will not repeat the entire
proof, but we only highlight some of the key points.  For more
information, the reader is referred to \cite{MPS98}.  

Let us consider, first of all, the set $\calM$ of morphisms $\phi: V
\times V \rightarrow G$ that solve the discrete field equations.  We
also assume that $\calM$ can be given the structure of a smooth,
infinite-dimensional manifold.  Then, a \emph{first variation} of an
element $\phi$ of $\calM$ is a section $\Gamma$ of $\prl$ such that
the associated finite variation transforms $\phi$ into new solutions
of the discrete field equations.  As the action sum $S$ can be
interpreted as a function on the set of morphisms from $V \times V$ to
$G$, and hence defines by restriction a function (also denoted by $S$)
on the set $\calM$.

By an argument similar to \cite[thm.~4.1]{MPS98}, it can then be shown
that, for any $\phi \in \calM$ and $\Gamma_1, \Gamma_2$ first
variations of $\phi$, the trivial identity $\d^2 S(\phi)(\Gamma_1,
\Gamma_2) \equiv 0$ can equivalently be written as 
\[
  0 = \sum_{[x] \cap \partial U \ne \varnothing} \left( \sum_{l; [x]_l
      \in \partial U} \Big( \Omega^{(l)}_L(\psi([x]))(\Gamma_1,
    \Gamma_2) \Big) \right).
\]
This characterization of multisymplecticity involves only the
quadrangles $[x]$ that contain edges which are part of the boundary
$\partial U$.

\section{Examples} \label{sec:examples}

\subsection{The pair groupoid $G = Q \times Q$} \label{sec:pair}

In this section, we treat in detail the case where $G$ is the pair
groupoid $Q \times Q$.  The results we obtain in this case agree with
those in \cite{MPS98}, which serves as a justification for our
approach.  

In this case, it is easy to see that $\bfG^k$ is just $Q^k$: the
identification is given by 
\[
  \big( (q_k, q_1), (q_1, q_2), \ldots, (q_{k-1}, q_k) \big) \mapsto
  (q_1, q_2, \ldots, q_k).
\]
Furthermore, the prolongation algebroid $\prl$ can be identified with
the $k$-fold Cartesian product of $TQ$ with itself.  For a vector
field $v$ on $Q$, the $i$th tangent lift of $v$ is the following
section of $(TQ)^k$: 
\[
  v^{(i)}: (q_1, q_2, \ldots, q_k) \mapsto (0, \ldots, 0, v(q_i), 0,
  \ldots, 0),
\]
where $v(q_i)$ occupies the $i$th place.

Now, let $L: \bfG^k \rightarrow \bfR$ be a Lagrangian and denote by
$\hat{L}$ the induced map on $Q^k$.  Then the $i$th Poincar\'e-Cartan
form is defined as 
\[
  \theta^{(i)}_L(q_1, \ldots, q_k; v_1, \ldots, v_k) = \d \hat{L}(q_1,
  \ldots, q_{i-1}, \cdot, q_{i+1}, \ldots, q_k) \cdot v_i,
\]
where $v_i \in T_{q_i} Q$ for $i = 1, \ldots, k$.  This was the
original definition of the Poincar\'e-Cartan forms in \cite{MPS98}.

It is instructive to see what becomes of the concepts of finite and
infinitesimal variations in this case: an infinitesimal variation is
just a vector field on $Q$, whereas a finite variation is the flow of
such a vector field.

As we pointed out before, a morphism $\phi: V \times V \rightarrow Q
\times Q$ can be seen as an assigment of an element of $Q$ to each
vertex in $V$.  For the case of the square mesh of
figure~\ref{fig:quadrangles}, we may therefore describe the field by
assigning a value $\phi_{i,j} \in Q$ to each vertex $(i,j)$.  Let
$\hat{L}(q_1, q_2, q_3, q_4)$ be a Lagrangian density; then
$\{\phi_{i,j}\}$ is a solution of the field equations (\ref{FE})
associated to $L$ if and only if, for all $(i, j) \in V$,
\begin{align*}
  &\frac{\partial L}{\partial q_1}(\phi_{i,j}, \phi_{i+1, j},
  \phi_{i+1, j+1}, \phi_{i,j+1})  +
  \frac{\partial L}{\partial q_2}(\phi_{i-1,j}, \phi_{i,j},
  \phi_{i,j+1}, \phi_{i-1, j+1})  + \\
  &\frac{\partial L}{\partial q_3}(\phi_{i-1,j-1}, \phi_{i,j-1},
  \phi_{i,j}, \phi_{i-1,j}) + 
  \frac{\partial L}{\partial q_4}(\phi_{i,j-1}, \phi_{i+1,j-1},
  \phi_{i+1, j}, \phi_{i,j})  = 0.
\end{align*}
These equations were first derived in \cite{MPS98}.

\subsection{The Lie-Poisson equations}

Consider now the case where the standard fibre $Q$ is a point, and the
groupoid $G$ a Lie group.  We will take a particular
\emph{triangular} mesh in $X = \bfR^2$, constructed as follows.  The
vertices are the points in $\bfR^2$ with integer coordinates:  
\[
  V = \{ (i, j) \in \bfR^2 : i, j \in \bfZ \}.
\]
However, instead of first specifying the set of edges, we start
from the set of faces, which we define to be 
\[
  F = \{ \big( (i, j), (i+1, j), (i+1, j+1) \big) \in V \times V
  \times V \}.
\]
The set of edges then consists of ``horizontal'' edges of the form
$((i, j), (i+1, j))$, ``vertical'' edges of the form $((i, j), (i,
j+1))$, and ``diagonal'' edges of the form $((i, j), (i+1, j+1))$.
This type of mesh was used in \cite{MPS98} as well.  The idea behind
it is that, in the appropriate physical setting, the horizontal edges
represent the spatial direction, whereas the vertical direction
represents the time direction.

Let us now consider a discrete Lagrangian $L: \bfG^3 \rightarrow
\bfR$.  Note that $\bfG^3$ is diffeomorphic to $G \times G$ by mapping
an element $(g_1, g_2, g_3) \in \bfG^3$ to $(g_1, g_2)$.  We denote
the induced Lagrangian by $\hat{L}$, where $\hat{L}(g_1, g_2) = L(g_1,
g_2, g_3)$.  Given the fact that our triangular mesh is different from
the ones we used in the body of the text, it is perhaps useful to
derive the field equations from scratch.  Let $\phi: V \times V
\rightarrow G$ be a morphism and consider a variation $\Psi$ of $\phi$
over some finite domain $\hat{U}$ in $\bfG^3$.  It is easy to see
that, if $[g] = \psi([x])$ is an element of $\bfG^3$, then the effect
of $\Psi$ on $[g]$ is as follows:
\[
   [g] \mapsto (h_1(t)^{-1}g_1h_2(t), h_2(t)^{-1}g_2 h_3(t),
   h_3(t)^{-1}g_3 h_1(t)),
\]
where $h_1, h_2, h_3$ are now \emph{arbitrary} curves in $G$, such
that $h_1(0) = h_2(0) = h_3(0) = e$, the unit in $G$.

\begin{figure}
\begin{center}
  \includegraphics[scale=0.7]{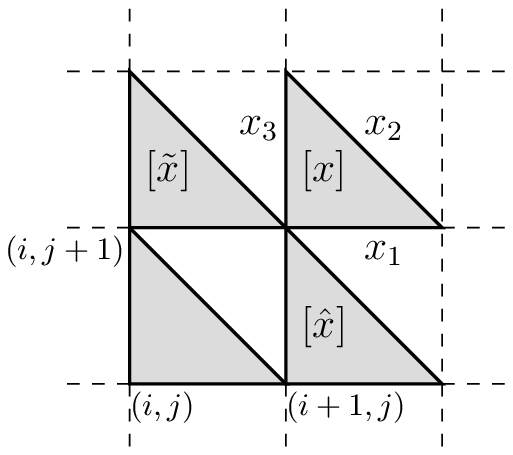}
  \caption{Triangular mesh in $\bfR^2$.}
  \label{fig:triangles}
\end{center}
\end{figure}

Let us now focus on the factor $h_1(t)$.  Following essentially the same
reasoning as in remark~\ref{remark:composable}, we see that $h_1(t)$
appears not only in the variation of $[g]$ but in the variation of two
additional triangles $[\hat{g}] = \psi([\hat{x}])$ and $[\tilde{g}] =
\psi([\tilde{x}])$ in the image of $\psi$ as well (see
figure~\ref{fig:triangles}).

The terms in the action sum involving $h_1(t)$ are therefore
\begin{align*}
  S(\phi_t) = \cdots & + {L}(h_1(t)^{-1}g_1h_2(t),
  h_2(t)^{-1}g_2h_3(t), h_3(t)^{-1}g_3h_1(t)) \\
    & + {L}( k_1(t)^{-1}\hat{g}_1k_2(t),
    k_2(t)^{-1}\hat{g}_2h_1(t), h_1(t)^{-1}\hat{g}_3k_1(t)) \\
    & + {L}( m_1(t)^{-1}\tilde{g}_1h_1(t),
    h_1(t)^{-1}\tilde{g}_2m_2(t), m_2(t)^{-1}\tilde{g}_3m_1(t)) \\ 
    & + \cdots,
\end{align*}
where $k_1(t), k_2(t)$ and $m_1(t), m_2(t)$ correspond to the effect
of $\Psi$ on the other vertices.  We now rewrite this in terms of the
induced Lagrangian $\hat{L}$ and demand that $\phi$ extremizes the
action sum to obtain the following set of discrete field equations:
\begin{align*}
  0 = \ddt S(\phi_t) & = \d \big[ \hat{L}(\cdot, g_3) \circ r_{g_1}
  \circ i\big] \cdot v_1 + \d \big[ \hat{L}(g_1, \cdot) \circ l_{g_3}
  \big] \cdot v_1 \\
  & + \d \big[ \hat{L}(\hat{g}_1, \cdot) \circ r_{g_3} \circ i\big]
  \cdot v_1 + \d \big[ \hat{L}(\cdot, \tilde{g}_3) \circ
  l_{\tilde{g}_1} \big] \cdot v_1,
\end{align*}
where $v_1 = \dot{h}_1(0)$.  As $h_1(t)$ is arbitrary, this implies
that, for any six elements in the image of $\phi$, distributed as in
figure~\ref{fig:triangles}, the following discrete field equations
must hold:
\[
\Big(l_{g_3}^\ast \d \hat{L}(g_1, \cdot) - r_{\hat{g}_3}^\ast \d
\hat{L}(\hat{g}_1, \cdot) \Big) + \Big(l_{\tilde{g}_1}^\ast
\d\hat{L}(\cdot, \tilde{g}_3) - r_{g_1}^\ast \d \hat{L} (\cdot, g_3)
\Big) = 0.
\]
In this expression, one can recognise, roughly speaking, two separate
discrete Lie-Poisson equations (see \cite{MarsdenPekarskyShkoller99}),
one for the ``spatial'' direction and one for the ``time'' direction.

The discrete Lie-Poisson equations arise, among others, in the context
of reduction.  Let $G$ be a Lie group and consider the pair groupoid
$G \times G$ over $G$.  Now, if $L: G^k \rightarrow \bfR$ is a
discrete Lagrangian, which is \emph{left invariant} in the sense
that $L(gh_1, gh_2, gh_3) = L(h_1, h_2, h_3)$ for all $g, h_1, h_2,
h_3 \in G$, and we consider the induced Lagrangian $L': G^{k-1}
\rightarrow \bfR$ defined as
\[
  L'(h_1^{-1}h_2, h_2^{-1}h_3, \ldots, h_{k-1}^{-1}h_k) = L(h_1, h_2,
  \ldots, h_k),
\]
then the following holds: a discrete field $\phi: V \times V
\rightarrow G$ will solve the field equations for $L$ if its reduced
field $\phi: V \times V \rightarrow G$ solves the Lie-Poisson
equations.  This is proved below in greater generality.  Note that one
recovers the Lie-Poisson equations by considering the morphism $\Phi:
(g, h) \mapsto g^{-1}h$.

\begin{theorem}
  Let $G'$ be a Lie groupoid over a manifold $Q'$ and consider a
  morphism $(\Phi, f): (G, Q) \rightarrow (G', Q')$.  Furthermore, let
  $L': \bfG'^k \rightarrow \bfR$ be a Lagrangian on $\bfG'^k$ and
  consider the induced Lagrangian $L = L' \circ \Psi$ on $\bfG^k$,
  where $\Psi: \bfG^k \rightarrow \bfG'^k$ is the map associated to
  $\Phi$.  

  A morphism $\phi : V \times V \rightarrow G$ will satisfy the
  discrete field equations for $L$ if the induced morphism $\Phi \circ
  \phi: V \times V \rightarrow G'$ satisfies the field equations for
  $L'$. 
\end{theorem}
\begin{mproof}
The proof relies on the following equality: for $i \le k$, $[g]
\in \bfG^k$, and $v \in A_{x}G$, where $x = \alpha(g_i)$,
\[
    v^{(i)}_{[g]}(L) = \left[ A\Phi(v) \right]^{(i)}_{\Psi([g])}(L'),
\]
which is relatively straightforward to prove.

With the same notations as above, this implies that 
\[
  \mathcal{EL}([g], [\hat{g}], [\tilde{g}]) \cdot v =
  \mathcal{EL'}(\Psi([g]), \Psi([\hat{g}]), \Psi([\tilde{g}])) \cdot
  (A\Phi(v)),
\]
where we have defined the Euler-Lagrange operator $\mathcal{EL} :
\bfG^3 \times \bfG^3 \times \bfG^3 \rightarrow A^\ast G$ as
\[
  \mathcal{EL}([g], [\hat{g}], [\tilde{g}]) \cdot v = 
    v^{(1)}_{[g]}(L) + v^{(2)}_{[\tilde{g}]}(L) + v^{(3)}_{[\hat{g}]}(L).
\]
Therefore, if $\phi$ is such that $\Phi \circ \phi$ is a solution of
the Euler-Lagrange equations for $L'$, then $\phi$ itself is a
solution of the Euler-Lagrange equations for $L$.
\end{mproof}

Lie-Poisson reduction in discrete field theories is thus very similar
to the corresponding theory in mechanics.  We glossed over some subtle
differences, however, mainly related to the reconstruction problem.
This will be treated in more detail in a forthcoming paper.

\subsection{Lattice gauge theories and discrete connections}

The geometrical setup described in section~\ref{sec:discretebundle} is
very similar to the one used in the treatment of gauge fields on a
lattice (see e.g. \cite{lattice} and the references therein).  

Let us consider an arbitrary compact Lie group $G$, which we interpret
as a Lie groupoid over a singleton $\{e\}$, with $e$ the unit element
of $G$.  For definiteness, we assume that the base space $X$ is once
again $\bfR^2$ and that a triangulation of $X$ is given.

A \emph{discrete gauge field} or \emph{discrete connection} is a map
$\psi: E \rightarrow G$, assigning a group element to each edge in
$\bfR^2$.  The \emph{field strength} or \emph{curvature} of such a
gauge field is the map $\Omega: F \rightarrow G$ which assigns to each
face $f$ the product
\[
   \Omega(f) = \psi(e_1) \cdot \psi(e_2) \cdot \psi(e_3),
\]
where $e_1$, $e_2$ and $e_3$ are the edges of $f$.  Here, we tacitly
assume that the edges are oriented and that $\psi(e^{-1}) =
\psi(e)^{-1}$.

Interpreting the gauge group $G$ as a Lie groupoid over $e$, it is
obvious that discrete fields, in the sense of definition~\ref{def:DJ},
correspond to flat gauge fields (i.e. gauge fields with vanishing
field strength).  Indeed, it is precisely the fact that these discrete
fields are groupoid morphisms, that makes the field strength vanish.

However, much of our formalism can be extended to the case of
arbitrary, non-flat gauge fields.  Indeed, let us consider a gauge
field $\psi: E \rightarrow G$.  In \cite{Baez}, the authors consider a
groupoid $\calP$, the units of which are the elements of $V$, while
the elements of $\calP$ are paths in $E$, i.e. sequences of composable
elements $e_1, e_2, \ldots, e_m$ in $E$.  A gauge field then gives
rise to a morphism $A: \calP \rightarrow G$ as follows:
\[
  A: (e_1, e_2, \ldots, e_m) \mapsto
  \psi(e_1)\psi(e_2)\cdots\psi(e_m).  
\]

If $A$ maps closed loops in $E$ to the unit in $G$, the associated
gauge field is flat.  In this case, we have the following interesting
property: simplicially homotopic paths are mapped to the same element
in $G$.  This is the discrete version of a well-known property of
continuous connections: if $\omega$ is a flat connection, and $\gamma,
\gamma'$ are closed loops that are homotopic, then the holonomy of
$\gamma$ is equal to that of $\gamma'$ (see~\cite[p.~93]{KN1}).  We
therefore obtain a morphism $\hat{A}$ from $\hat{\calP}$, the groupoid
of paths modulo simplicial homotopy, to $G$.  In the case where $X =
\bfR^2$, $\hat{\calP}$ can be identified with $V \times V$, and we are
back at our starting point, that of representing flat gauge fields by
morphisms from $V \times V$ to $G$.

\begin{remark}
  At first, our use of flat discrete connections might seem to exclude
  the treatment of gauge theories.  A closer look will reveal that the
  flatness used in the main body of our text plays a similar role as
  the flatness (or integrability) of the connections used in the
  connection-theoretic De Donder-Weyl treatment of classical field
  theories (see \cite{CrampinMS, Saunders89}) and is hence quite
  unrelated to the curvature of the fields.
\end{remark}

\section{Conclusions and outlook}

In this paper, we have described a geometric model for discrete field
theories.  We extended the foundational work done in \cite{MPS98} by
allowing for discrete fields that take values in an arbitrary groupoid
and we showed that much of the geometric structures from (discrete)
field theory, such as the Poincar\'e-Cartan forms and the notion of
multisymplecticity, carry over quite naturally to this setup. 

There remain many interesting open problems in this area.  In a future
publication, we intend to investigate the problem of discrete
reduction into further detail, as well as the reconstruction problem.
It turns out that, just as in the continuous case (see~\cite{MarcoLMP,
  MarcoRatiu, MarcoAMS}) there appears an additional condition involving
discrete curvature, which is absent from the reconstruction problem in
mechanics. 

Another interesting link concerns the theory of discrete integrable
fields as proposed by Bobenko, Suris, and coworkers (see
\cite{Adler01, BSbook, BSquadgraph, BobenkoPinkall} as well as the
references therein).  After all, the kind of fields that we
investigate here bear some tantalizing resemblances to their
zero-curvature representations.

Ultimately, and perhaps not unrelated to the previous point, one would
hope that the techniques developed in this paper can be applied to the
construction of robust integrators for PDEs.  It should be stressed,
however, that the concept of ``symplectic integrator'' for field
theories is much more subtle than for simple mechanical systems and
that the issue whether multisymplectic integration schemes provide
qualitatively better results is usually decided on a case-by-case
basis (see for instance \cite{AscherMcLachlan}, where a number of
symplectic and multisymplectic schemes are compared in the case of the
celebrated Korteweg-de Vries equation).

\section*{Acknowledgements} 

Financial support of the Research Foundation--Flanders
(FWO-Vlaanderen) is gratefully acknowledged.  Furthermore, we would
like to thank Bavo Langerock, David Mart\'{\i}n de Diego and Eduardo
Mart\'{\i}nez for many stimulating discussions.

\bibliographystyle{ams-abbrv-jvk}
\bibliography{biblio}

\end{document}